\documentclass[fleqn,usenatbib]{mnras}

\usepackage{newtxtext,newtxmath}

\usepackage[T1]{fontenc}
\usepackage{ae,aecompl}

\usepackage{graphicx}	
\usepackage{amsmath}	
\usepackage{amssymb}	
\usepackage{float}
\usepackage{subfig}
\usepackage{multirow}

\outer\def\gtae {$\buildrel {\lower3pt\hbox{$>$}} \over 
{\lower2pt\hbox{$\sim$}} $}
\outer\def\ltae {$\buildrel {\lower3pt\hbox{$<$}} \over 
{\lower2pt\hbox{$\sim$}} $}
\newcommand{\Msun}{$M_{\odot}$}

\newcommand{\Rsun}{$R_{\odot}$}

\newcommand{\kep}{{\it Kepler}}

\newcommand{\tess}{{\it TESS}}

\title[Variability in Solar-Type Stars]{Superflares and Variability in Solar-Type Stars with \tess\ in the Southern Hemisphere}

\author[L. Doyle et al.]{
L. Doyle,$^{1,2}$\thanks{E-mail: lauren.doyle@armagh.ac.uk}
G. Ramsay,$^{1}$
J. G. Doyle,$^{1}$ 
\\
$^{1}$Armagh Observatory and Planetarium, College Hill, Armagh, BT61 9DG\\
$^{2}$Mathematics, Physics and Electrical Engineering, Northumbria University, Newcastle upon Tyne, NE1 8ST
}

\date{Accepted 2020 March 30. Received 2020 March 30; in original form 2020 January 14}

\pubyear{2020}

\begin{document}
\label{firstpage}
\pagerange{\pageref{firstpage}--\pageref{lastpage}}
\maketitle

\begin{abstract}
Superflares on solar-type stars has been a rapidly developing field ever since the launch of \kep. Over the years, there have been several studies investigating the statistics of these explosive events. In this study, we present a statistical analysis of stellar flares on solar-type stars made using photometric data in 2-min cadence from \tess\ of the whole southern hemisphere (sectors 1 -- 13). We derive rotational periods for all stars in our sample from rotational modulations present in the lightcurve as a result of large starspot(s) on the surface. We identify 1980 stellar flares from 209 solar-type stars with energies in the range of $10^{31} - 10^{36}$~erg (using the solar flare classification, this corresponds to X1 - X100,000) and conduct an analysis into their properties. We investigate the rotational phase of the flares and find no preference for any phase suggesting the flares are randomly distributed. As a benchmark, we use GOES data of solar flares to detail the close relationship between solar flares and sunspots. In addition, we also calculate approximate spot areas for each of our stars and compare this to flare number, rotational phase and flare energy. Additionally, two of our stars were observed in the continuous viewing zone with lightcurves spanning one year, as a result we examine the stellar variability of these stars in more detail. 
\end{abstract}




\section{Introduction}
Solar flares are powerful, eruptive events which are seen across the entire electromagnetic spectrum. They result from a build up of magnetic energy which is converted into kinetic energy, thermal energy and particle acceleration through magnetic reconnection, where the magnetic field simplifies. Overall, our Sun can show flares with energy outputs ranging from $10^{24}$ - $10^{32}$~erg \citep{aschwanden2000time}. However, studies of solar-type stars using \kep\ have revealed flares with energies exceeding $10^{32}$~erg, with `superflares' having energies up to $10^{38}$~erg \citep{schaefer2000superflares}.  

During its lifetime (2009 -- 2018), the \kep\ mission \citep{borucki2010kepler} provided a wealth of photometric lightcurves for hundreds of thousands of stars. This allowed for the study of stellar activity from the interior to the photosphere in a large selection of stars from solar-type to low mass. 

\cite{maehara2012superflares} conducted the first statistical study of flares on solar-type (G-type main sequence) stars. They used \kep\ long cadence (LC: 30-min) lightcurves and identified 365 superflares (flares with energies $> 10^{33}$~erg) on 148 G-type stars. In addition, they fit the occurrence distribution rate of the flares  with a power law and find it is similar to solar flares and flares on low mass stars. They also explore the proposed theory of hot Jupiters being important in the generation of superflares \citep{rubenstein2000superflares} and find none have been discovered around their sample of solar-type stars indicating they are rare. Lastly, using derived rotation periods for each star they conclude superflares occur more frequently on young solar-type stars (younger than our Sun) as a result of faster rotation periods. 

In \cite{shibayama2013superflares}, they extend the work started by \cite{maehara2012superflares}, searching for superflares on solar-type stars (G-type dwarfs) with \kep\ LC data over a longer period of 500~days. This resulted in identifying 1547 superflares on 279 solar-type stars, increasing the sample of flares by a factor of four. Overall, they confirm the previous results, identifying the distribution of occurrence rate as a function of energy to be similar to that of solar flares. Interestingly, by monitoring the brightness variation of their sample they conclude the high occurrence of superflares could be a result of extremely large starspots. 

More recently, \cite{notsu2019kepler} presented a complete review of \kep\ solar-type superflares including updates on a new sample using the Apache Point Observatory (APO) and Gaia DR2. The results from Gaia DR2 revealed the possibility of contamination of subgiant stars within the classification of \kep\ solar-type stars. This is due to previous classifications using $T_{eff}$ and $log(g)$ values from the \kep\ Input Catalog (KIC: \cite{brown2011kepler}) where there are large differences between real and catalogued values. One of the other key differences was their ability to check the binarity of their new sources using APO spectroscopic observations, ruling out stars which were members of binary systems. This in turn rules out the generation of flares as a result of magnetic interaction between the binary system. They also investigate starspot size, concluding the majority of superflares occur on stars with larger starspots, however, they acknowledge there is some scatter. With regards to rotation period ($P_{rot}$), they note maximum spot size does not depend on $P_{rot}$ but maximum flare energy does continuously decrease with slower rotation.

In solar physics, the relationship between flares and sunspots has been well established, with these phenomena being closely linked. Multiple studies, such as \cite{maehara2012superflares, notsu2013superflares, maehara2017starspot, notsu2019kepler}, of solar-type stars report close links between starspots and flaring activity concluding superflares are a result of stored magnetic energy near starspots. Despite this the relationship between flare rotational phase and starspots in solar-type stars has not been investigated in great detail. Large variations in brightness seen in lightcurves, also known as rotational modulation, are attributed to be the result of starspots on the stellar disk moving in and out of view as the star rotates \citep{rodono1986rotational, olah1997time}. If superflares do occur near starspots due to the storage of magnetic energy then you would expect to see a correlation between starspots and flare occurrence. 

In our previous studies, \cite{doyle2018investigating, doyle2019tess} (henceforth Paper I \& II respectively), we used K2 \& \tess\ short cadence (SC) photometric data to investigate the rotational phase of flares in a sample of 183 M dwarfs. By using simple statistical tests we determined the phase distribution of the flares was random and did not coincide with the large starspot producing the rotational modulation. This result came as a surprise as it suggests the flares on these M dwarf stars are not correlated with the dominant large starspot present on the stellar disk. As a result, this indicates the magnetic field and resulting activity on these stars may be more complex than what is observed on the Sun. 

In this study, we use \tess\ 2-min photometric lightcurves from a sample of solar-type stars observed in sectors 1 - 13 to conduct a statistical analysis of their flaring properties where the short cadence 2-min \tess\ data is important for detecting low energy, short duration flares. In addition to investigating the rotation periods, flare energies and flare frequency, we will explore the rotational phase of the flares. This analysis aims to determine whether the flares and starspots on these solar-type stars share the same strong correlation as solar flares and sunspots on the Sun. Furthermore, we will use historic GOES data of solar flares to investigate the relationship between solar flares and sunspots in greater detail. 

\begin{figure}
    \centering
    \includegraphics[width = 0.47\textwidth]{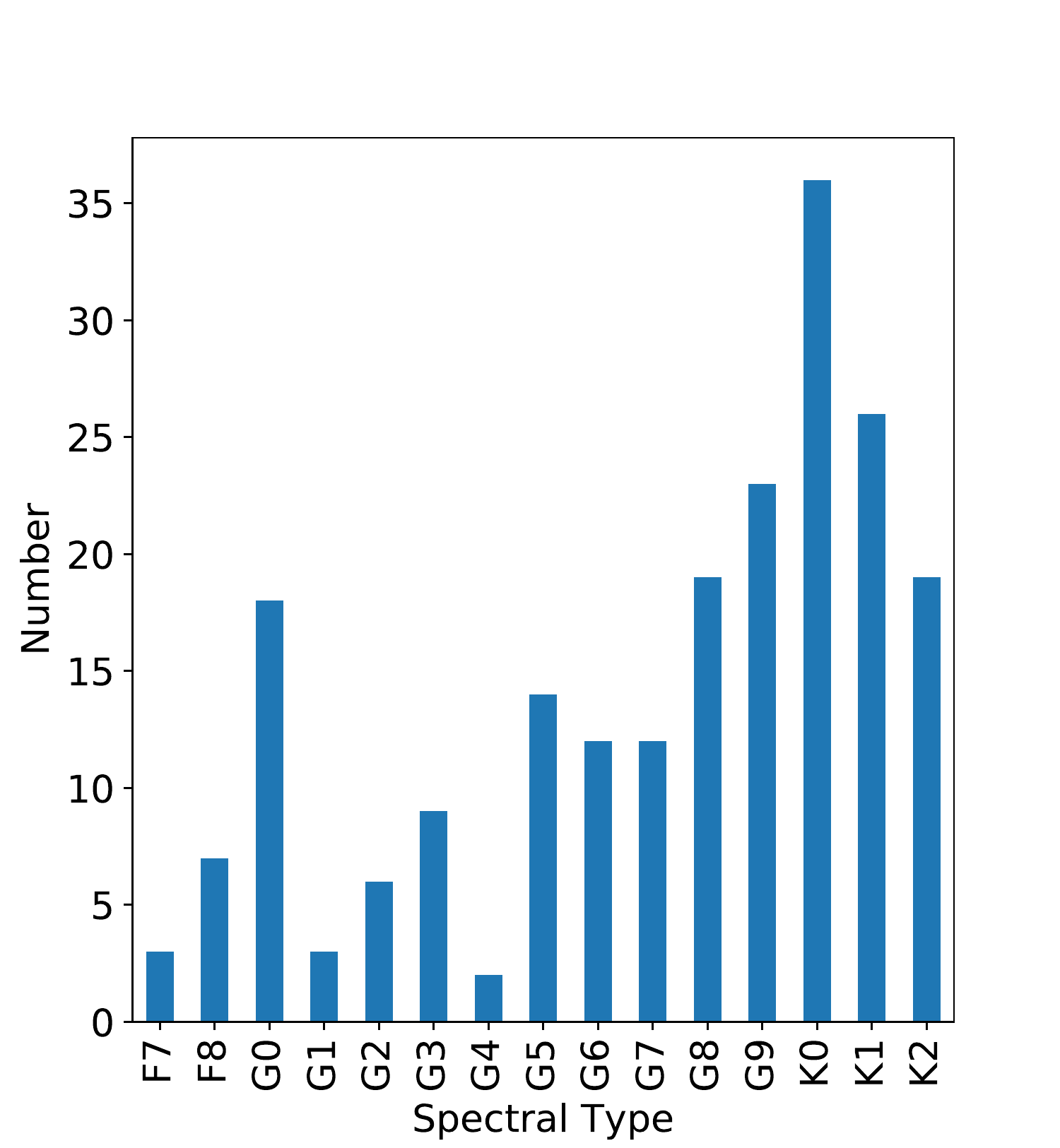}
    \caption{A histogram showing the spread in spectral types within our solar-type star sample observed in 2-min cadence by \tess.}
    \label{spectral_spread}
\end{figure}

\section{Solar-Type star sample}
\label{sample}

In previous solar-type star studies, the sources were identified using their effective temperatures ($T_{eff}$) and $log(g)$ values from the KIC, or other associated catalogues. However, this process led to a contamination of sub-giants within the sample which were incorrectly identified due to differences between real (Gaia DR2) and catalogued values. Here we identify solar-type stars as those with spectral types ranging from F7 - K2 according to the {\tt SIMBAD} catalogue\footnote{\url{http://simbad.u-strasbg.fr/simbad/}}, and have been observed in 2-min cadence by \tess.  

We now go on to discuss the various steps which were taken to eliminate any sources which were not main sequence solar-type stars. Firstly, the sources were cross referenced with the SkyMapper Southern Sky Survey \citep{wolf2018skymapper} and Gaia DR2 \citep{gaia18}. Any star which did not possess SkyMapper magnitudes or Gaia parallaxes was not considered any further. We use radii and luminosity values from Gaia DR2 to eliminate any which are likely to be giants and hence wrongly classified within {\tt SIMBAD}. We use Skymapper multi-colour magnitudes and Gaia parallaxes to determine the quiescent luminosity of the stars in the \tess\ bandpass. The magnitudes in the $g, r, i$ and $z$ bands are converted to flux and then fitted by a polynomial to produce template spectra of each star. These are then convolved by the \tess\ bandpass providing the quiescent flux of each star in the \tess\ bandpass. By inverting the Gaia parallaxes we determine the distances of our sample and use them to infer the quiescent luminosity of each star. The Gaia parallaxes for our sample are all much larger than their errors so, we do not expect spurious solutions as outlined in \cite{arenou2018gaia}. These values along with the stellar properties of each star are provided in Table \ref{stellar_properties}. 

\begin{figure*}
    \centering
    \includegraphics[width = 0.95\textwidth]{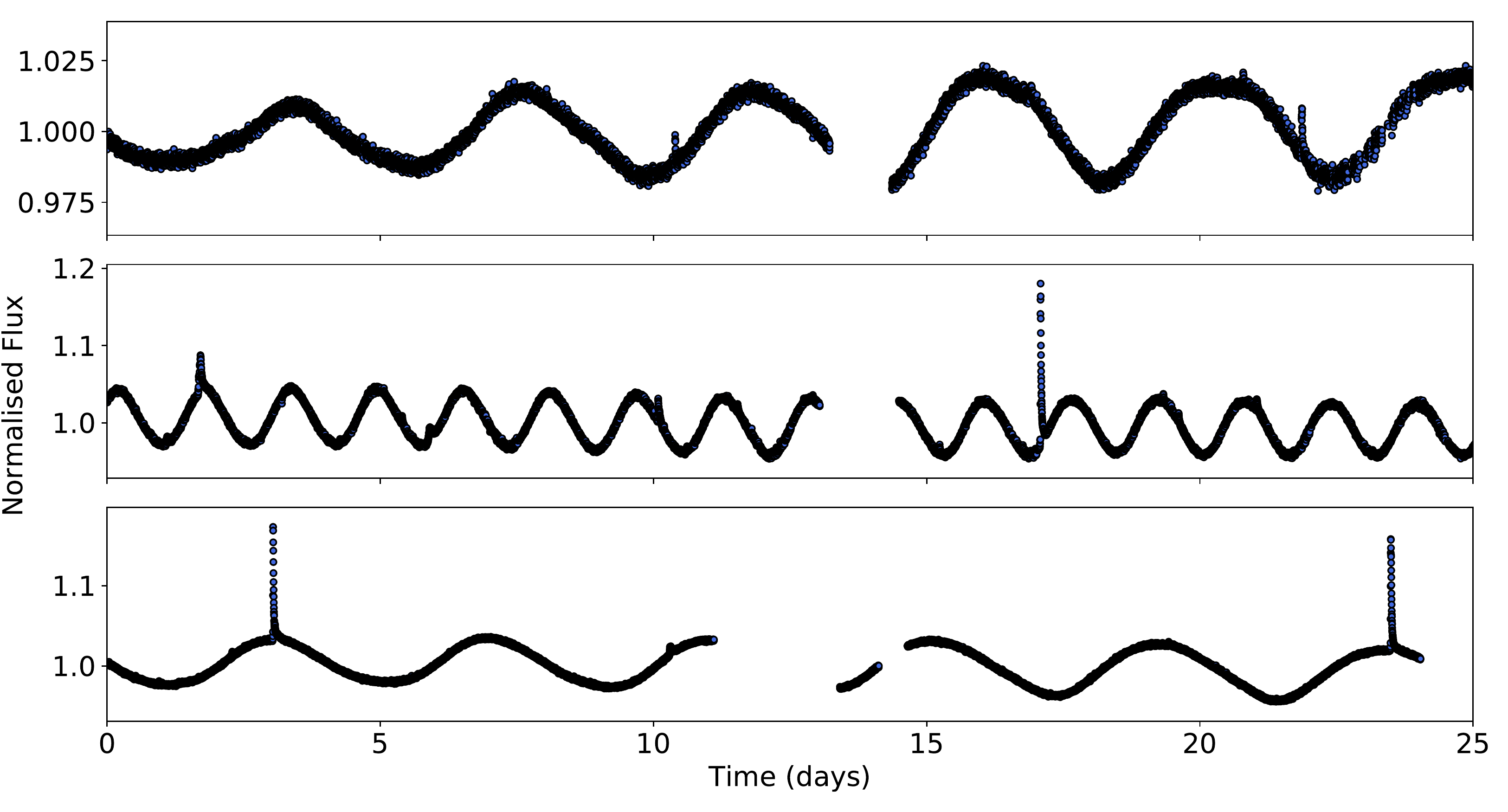}
    \caption{Here we show three examples of lightcurves from the solar-type stars CD-52 10232 (top), HD 221224 (middle) and BD-19 3018 (bottom). These have spectral types K0, G5 and K0 and rotation periods 4.25 days, 1.58 days and 4.07 days respectively. As well as clear modulation within these lightcurves as a result of starspots, flares can also be seen in all three stars. }
    \label{lightcurves}
\end{figure*}

The \tess\ 2-min lightcurves of the remaining sources were visually inspected by eye individually to determine those which showed any signs of rotational modulation. Some sources which showed complex lightcurves or no evidence of rotational modulation were not considered further. Only a handful of these sources which did not possess rotational modulation showed any evidence of flaring activity and were omitted as the rotational period is a key aspect of our analysis. Finally, all lightcurves were then run through a flare finding algorithm (see Section \ref{findflares} for more details) and any which did not show any flaring activity were omitted from further analysis. This process resulted in a final sample of 209 solar-type stars observed in 2-min cadence by \tess. The spread of spectral types within this sample is shown in Figure \ref{spectral_spread}. Additionally, it is important to note three of the stars within our sample show evidence of belonging to an eclipsing binary in the \tess\ lightcurves. We will discuss this further in Section \ref{discuss}. 

Due to the nature of the selection of targets for our solar-type sample there are some selection biases which should be highlighted. Firstly, only stars which had spectral types recorded in {\tt SIMBAD} were initially selected: we are therefore biased towards stars which had a spectral type recorded. Secondly only those stars which display rotational modulation are selected, as it is critical for our analysis. However, this results in a bias towards active solar-type stars in our sample meaning a complete picture is not achieved. Additionally, there is a bias towards later type (G8 - K2) solar-type stars, see Figure \ref{spectral_spread}. Finally, targets had to possess SkyMapper and Gaia data which resulted in some targets being omitted from this study which potentially could have shown rotational modulation and/or flares. Despite these selection biases, we consider that our final sample of 209 solar-type stars with spectral types between F7 and K2 (for the full stellar properties see Table \ref{stellar_properties}) is large enough for determining the rate of super-flares from solar-type stars and whether they show any rotational phase dependence.

\section{TESS Photometric Lightcurves}
\label{tess_lightcurves}
The Transiting Exoplanet Survey Satellite (\tess) \citep{Ricker2015} was launched in April 2018. During its initial two year mission, \tess\ will make a near all-sky survey observing 200,000 of the closest stars to our Sun. \tess\ is fitted with four CCD cameras which act as a $1 \times\ 4$ array providing a total FOV of $24\deg \times\ 96\deg$. Each of the hemispheres are split into 13 equal sectors which \tess\ will observe for a duration of 27.4 days each. Close to the poles there are overlaps within sectors producing certain areas which are observed in excess of $\sim 300$~days, this includes a continuous viewing zone. \tess\ data releases include Full Frame Images (FFIs) and Short Cadence (2-min) lightcurves. For this study we will be using the 2-min lightcurves as they allow for the observation of the shorter and typically less energetic flares.

The primary mission for \tess\ is to search for exoplanets via the transit method around low mass M dwarf stars. Therefore, \tess\ has a bandpass of 600 - 1000~nm and is centred on 786.5nm \citep{Ricker2015} making it is slightly redder in comparison to \kep. This redder bandpass allows \tess\ to observe a larger number of M dwarfs making planets easier to detect. As mentioned in our paper \cite{doyle2019tess}, the \tess\ bandpass is more sensitive to redder wavelengths meaning it will not detect less energetic flares as they typically have their peak emission towards the blue. Overall, the rms of the \tess\ lightcurves for M dwarfs of the same magnitude is 4.6 times lower than the \kep\ lightcurves. As a result, \tess\ will be unable to detect short duration, low amplitude flares \citep{ramsay2020ufr}, however, this is related to the brightness of the stars where lower energy flares could be detected on brighter targets. Overall, the key factor in detecting low energy flares is the SNR of the lightcurve.

In this paper we will use photometric lightcurves of solar-type stars from sectors 1 - 13 made between 25th July 2018 - 18th July 2019. We downloaded the calibrated lightcurve for each of our target stars from the MAST data archive\footnote{\url{https://archive.stsci.edu/tess/}}. We used the data values for {\tt PDCSAP\_FLUX}, which are the Simple Aperture Photometry values, {\tt SAP\_FLUX}, after the removal of systematic trends common to all stars in that Chip. Each photometric point is assigned a {\tt QUALITY} flag which indicates if the data may have been compromised to some degree by instrumental effects. We removed those points which did not have {\tt QUALITY=0} and normalised each lightcurve by dividing the flux of each point by the mean flux of the star.  Overall, 158 solar-type stars in the sample (76\%) were observed in only one sector with the remaining 58 (24\%) targets being observed in multiple sectors. The full details of how many sectors each star was observed in can be found in Table \ref{stellar_properties}.

\section{Stellar and Flare Properties}
In this section we will look at both the stellar and flare properties of each star. This includes determining the rotation period, identifying the flares and calculating their energies. In addition, we will also look further into a small group of ultra-fast rotators identified within our solar-type sample. 

\subsection{Rotation Period}
We determine rotation periods for all 209 solar-type stars in our sample using the rotational modulation observed in the lightcurve. This rotational modulation occurs as a result of large, dominant starspots which move in and out of view as the star rotates, changing the brightness of the star periodically. Examples of this phenomenon can be seen in Figure \ref{lightcurves} of several solar-type stars ranging in rotation period and spectral type. 

To determine the rotation periods, $P_{rot}$, we utilise a Lomb-Scargle (LS) periodogram from the software package {\tt vartools} \citep{hartman2016vartools}. This provides an initial estimation of $P_{rot}$, and by phase folding and binning the lightcurve through an iterative process a final value is verified. Along with $P_{rot}$ phase zero, $\phi_{0}$, is also determined and represents the minimum of the flux of the rotational modulation. Overall, this process allows for the determination of both $P_{rot}$ and $\phi_{0}$ which is used in subsequent analysis of the magnetic activity. Errors on $P_{rot}$ are estimated to be within a few percent. It is important to note that occasionally the LS periodogram detects half of the true period. However, as we have visually inspected each lightcurve we can usually identify instances where this has occurred and modify the period accordingly. We therefore do not consider that many stars have incorrect periods. 

The stellar properties of the sample can be seen in Table \ref{stellar_properties} including both $P_{rot}$ and $\phi_{0}$. Within our sample rotation periods range from 0.24 - 11.16 days. It is difficult to detect stars with $P_{rot}$ > 10 days (the Sun has $P_{rot} \sim$ 27 days), due to the observation length of \tess\ at $\sim$ 27 days per sector although for stars with more than one sector of data we can get longer periods. However, despite this, it is still possible to conduct an analysis on this sample in comparison to the Sun as we are comparing the relationship between flares and starspots.

\begin{table*}
\caption{The stellar properties of the first few stars in our survey detailing the rotation periods, quiescent luminosity, energy range and duration range of the flares. The apparent magnitude in the \tess\ band-pass, $T_{mag}$, is taken from the \tess\ Input Catalog (TIC) along with the TIC ID \citep{stassun2018tess}. The distances are derived from the Gaia Data Release 2 parallaxes \citep{gai16, gaia18} and the spectral types are obtained from the {\tt SIMBAD} catalogue. \newline
{\it This table is available in its entirety in a machine-readable form in the online journal. A portion is shown here for guidance regarding its form and content.}}

   \begin{center}
   \label{stellar_properties}
\resizebox{1.0\textwidth}{!}{
	\begin{tabular}{lccccccccccccc}
    \hline 
	Name    & TIC ID     &  sector      & Ra      & Dec     & No. of   & SpT  & $T_{mag}$ & Parallax  & Distance & $P_{rot}$ & $log(L_{star})$ & $log(E_{flare})$   &  Duration        \\
	        &            &              & (deg)   & (deg)   & Flares   &      &           & (mas)     & (pc)     & (days)    & (erg/s)         & (erg)              &  (minutes)       \\
	\hline

CD-52 10232  & 161172848  & 1   	    & 339.87  & -52.08  & 2        & K0   & 10.03     & 9.42      & 106.1   &  4.25     &  32.82          & 33.80 -- 34.26    &  34.0 -- 70.0 \\
HD 205297    & 403121294  & 1           & 324.31  & -65.03  & 11       & G6   & 8.27      & 13.64     & 73.3    &  1.52     & 33.05           & 33.09 -- 34.44    &  12.0 -- 102.0 \\
HD 49855     & 176873028  & 1,2 \& 4    & 100.94  & -71.97  & 12       & G6   & 8.39      & 17.98     & 55.6    &  3.85     & 32.78           & 31.73 -- 34.10    &  8.0 -- 52.0 \\    
HD 42270     & 261236136  & 1,12 \& 13  & 88.37   & -81.94  & 19       & K0   & 8.28      & 16.96     & 58.9    &  1.88     &  32.82          & 32.82 -- 35.06    & 8.0 -- 232.0 \\
    \hline
    \end{tabular}}
    \end{center}
\end{table*}

\subsection{Stellar Flares}
\label{findflares}

We use the same method as Papers 1 \& 2 to identify the flares present in each lightcurve and calculate their energies in the \tess\ bandpass. {\tt FBEYE} \citep{davenport2014kepler} is a suite of {\tt IDL} programs which scans the lightcurve, flagging any points which are over a $2.5\sigma$ threshold. Within the flare finding algorithm, potential flares are required to consist of two or more consecutive flagged points; all of the flares within our sample are composed of many more points and 2 is simply the minimum requirement. Additionally, all flares are visually checked by eye to validate they are indeed flares requiring  them to possess a classical flare shape with a sharp rise and exponential decay. Once complete, a comprehensive list of stellar flares including their start and stop times, flux peak and equivalent duration is produced for each star. Some examples of these flares can be seen in Figure \ref{lightcurves} along with the rotational modulation.

\begin{figure}
    \centering
    \subfloat{\includegraphics[width = 0.47\textwidth]{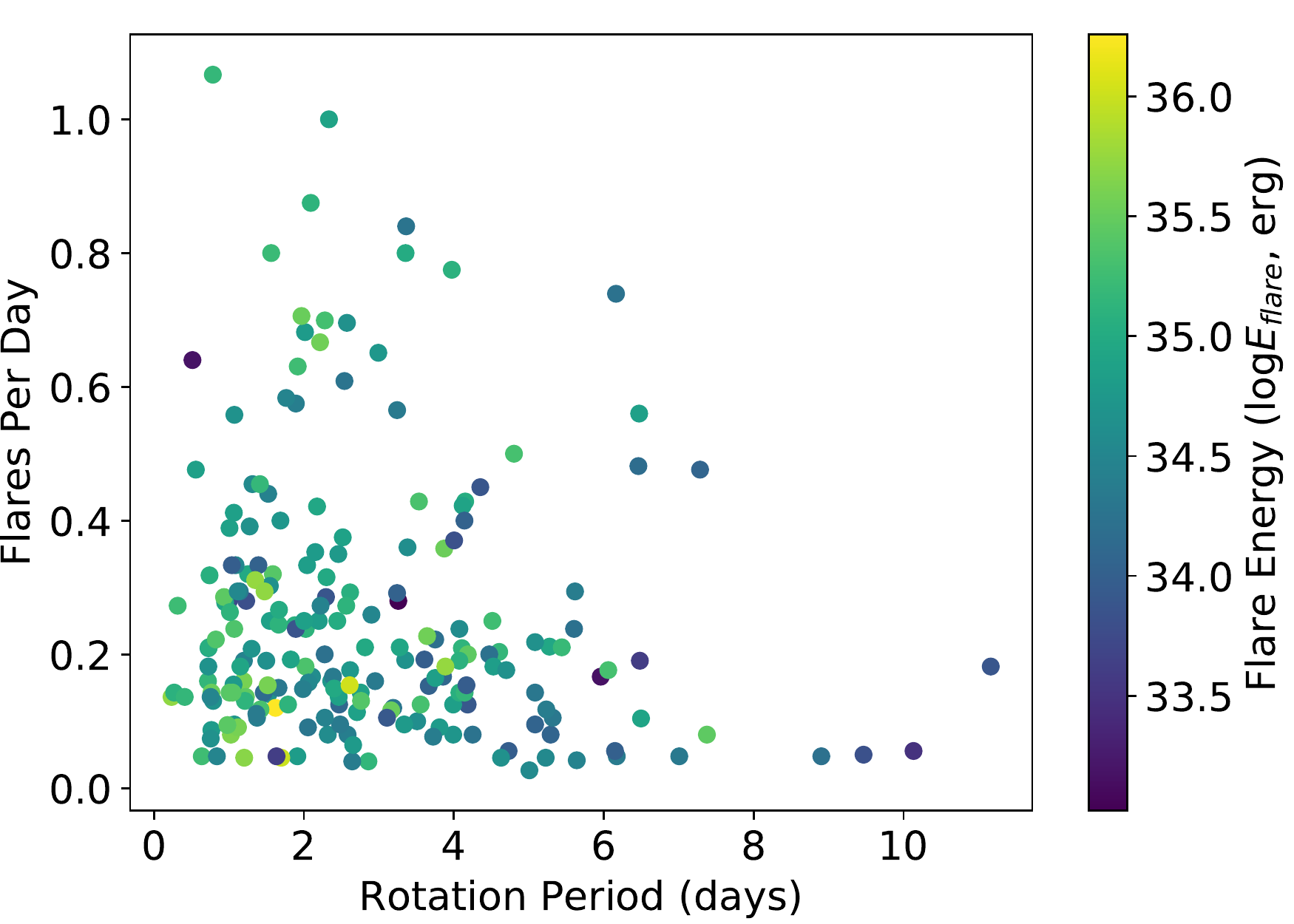}}\\
    \subfloat{\includegraphics[width = 0.47\textwidth]{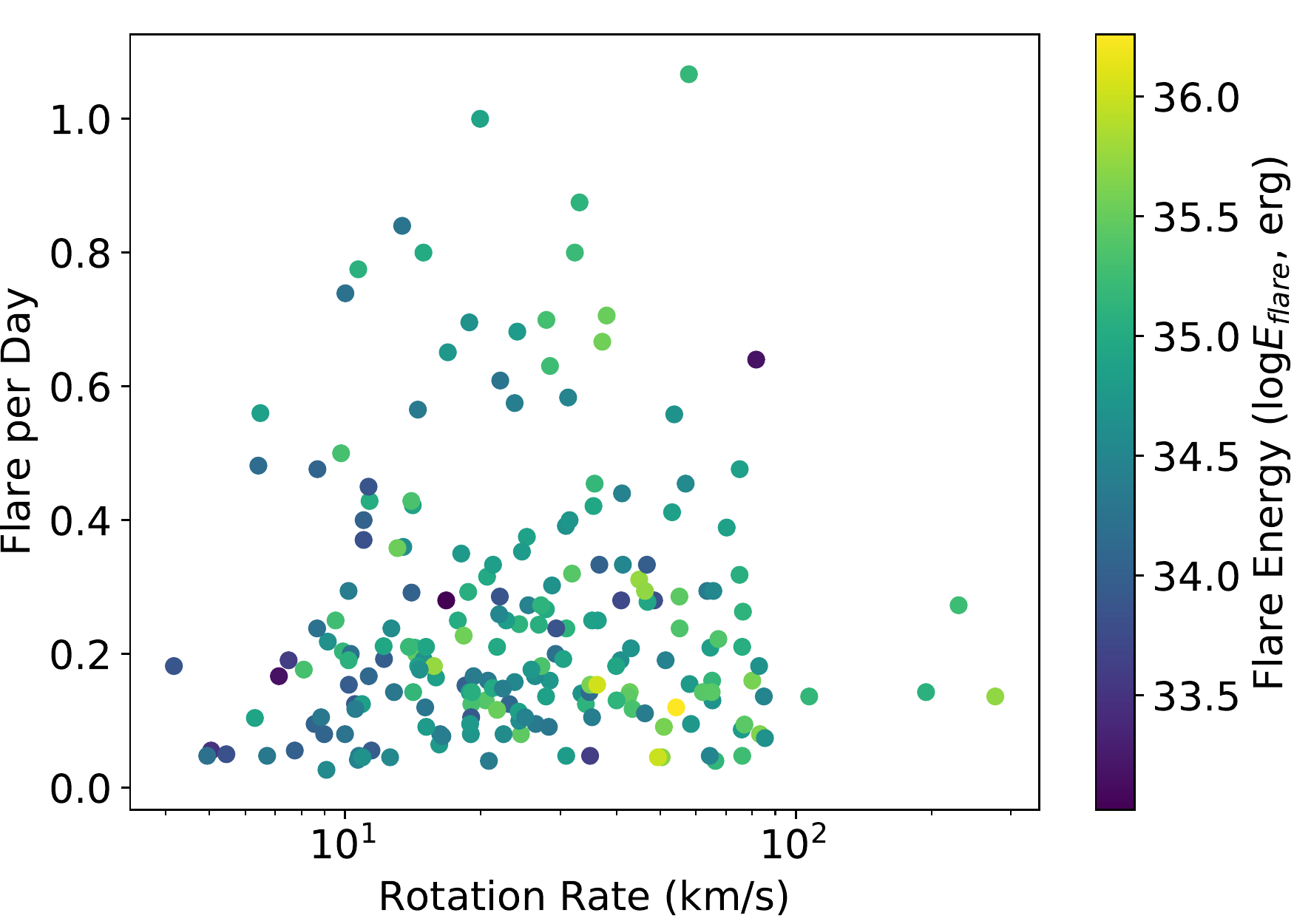}}
    \caption{The normalised flares per day of each star in our sample as a function of rotation period, $P_{rot}$, (top panel) and rotation rate , $\Omega$ (lower panel). Each of the points are colour coded according to the colour bar which represents the maximum flare energy from the star. }
    \label{rot_flareno}
\end{figure}

\begin{figure}
    \centering
    \includegraphics[width = 0.47\textwidth]{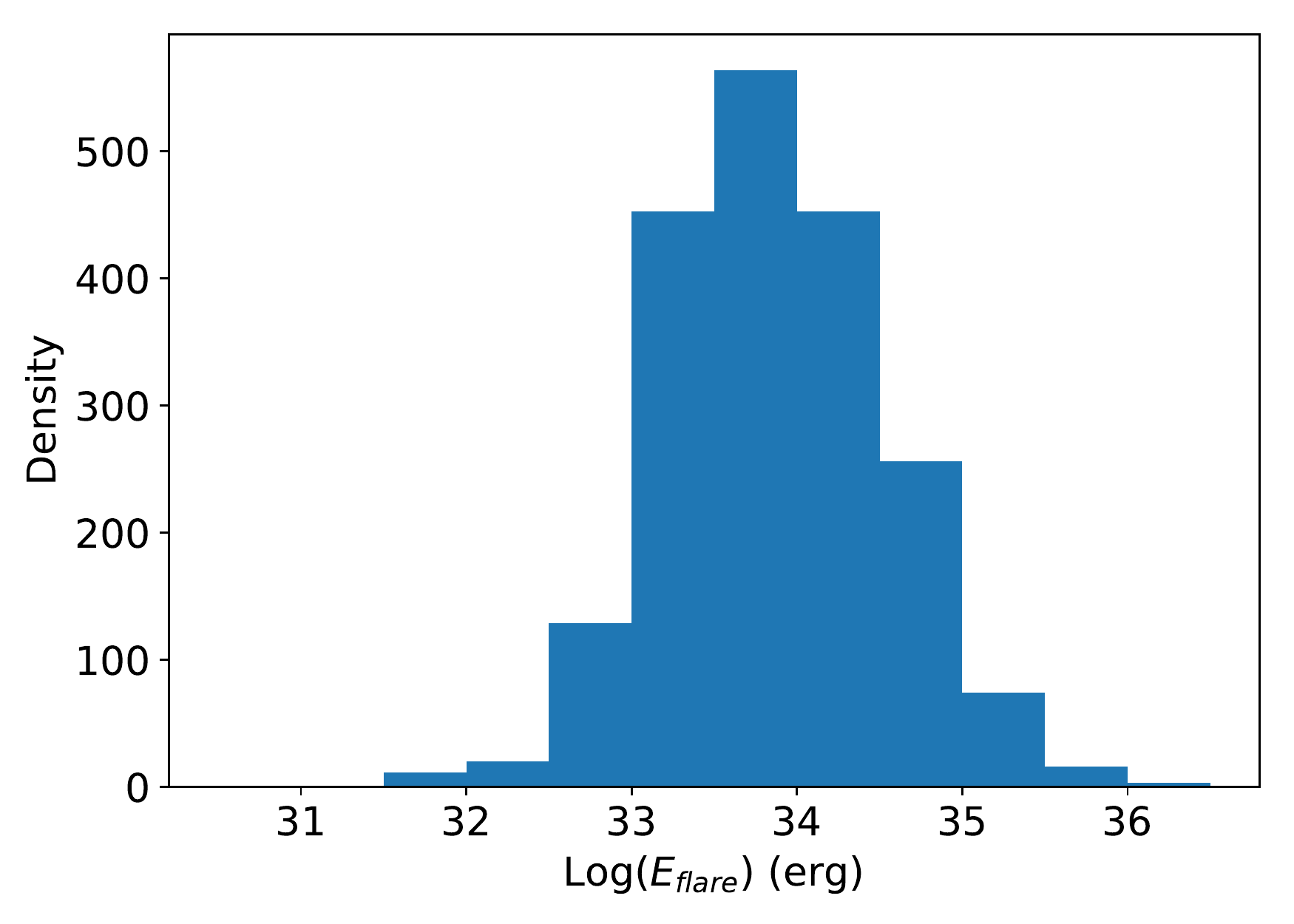}
    \caption{The distribution of the energies of the 1980 flares for our sample of 209 solar-type stars. This histogram was used to define the cut off between high and lower energy flares of $10^{34}$~erg which was used in subsequent analysis. }
    \label{hist_flares}
\end{figure}

The flare numbers for each star can be normalised to give the number of flares per day as the observation length of each star varies as a result of being observed in multiple sectors. In Figure \ref{rot_flareno} we plot the normalised flare number alongside the rotational period which shows flare number decreasing with increasing rotation period. Despite the lack of stars with $P_{rot}$ > 10 days, this is consistent with other studies such as \cite{stelzer2016rotation} but also with Papers 1 \& 2. 

In addition, we also plot the rotation rate, $\Omega$, as a function of the normalised flare number, see Figure \ref{rot_flareno}. To do this we use the relationship $\Omega = 2\pi R/P_{rot}$  where $R$ is the radius of the star taken from the Gaia DR2 release and $P_{rot}$ is the rotation period derived earlier. This allows us to identify the fast rotators within the sample. In Paper II, we discovered a group of M dwarf ultra-fast rotators (UFRs) with $P_{rot}$ < 0.3 days which surprisingly displayed a low level of flaring activity. Determining ages for these stars did not provide any explanation for their peculiar behaviour. In this sample we have four solar-type UFRs with $P_{rot}$ < 0.4 days which also show low levels of flaring activity. At the moment, we do not have a clear explanation for this phenomenon, however, we do believe it is related to the magnetic field properties of the stars. Since these objects are fast rotators, they are probably young, hence it is possible that the flares have their maximum energy in the blue, thus the \tess\ band-width only sees the more energetic events.

Next we want to determine the energies of the flares within the \tess\ bandpass. These are calculated as the equivalent duration, area under the flare lightcurve obtained from {\tt FBEYE}, multiplied by the quiescent stellar luminosity. As mentioned previously the quiescent stellar luminosity, $L_{star}$, was determined from both Skymapper magnitudes and Gaia parallaxes, full details in Section \ref{sample}. Within the solar-type sample a large variety of flare energies are seen ranging from $2.1 \times 10^{31} - 1.8 \times 10^{36}$~erg. Approximately 92\% of our flare sample are classified as superflares with energies greater than $10^{33}$~erg. It is important to note here the term `superflare' was defined according to the Carrington event which was a X45 class solar flare observed in 1859 and had an energy output of $4.5 \times 10^{32}$~erg \citep{cliver20131859}. Hence, the majority of our flares within the sample exceed this energy range making them `superflares'. Similarly, only 1.6\% of our flare sample have energies less than $10^{32}$~erg, which is the range of solar flares, with no flares less than $10^{31.5}$~erg. This lower limit is is due to the effective area of the TESS telescopes and is also related to the stellar brightness, see Section \ref{tess_lightcurves}. The highest energy flare of $1.8 \times 10^{36}$~erg was observed on the star HD 217344 (TIC 229066844) a G4 star with $P_{rot}$ of 1.62 days. All of the flare properties including the minimum flare energy detected (in the online version) for each star can be seen in Table \ref{stellar_properties}, and the spread of flare energies can be seen in Figure \ref{hist_flares}.

\subsection{Superflare Frequency}
\begin{figure}
    \centering
    \includegraphics[width = 0.47\textwidth]{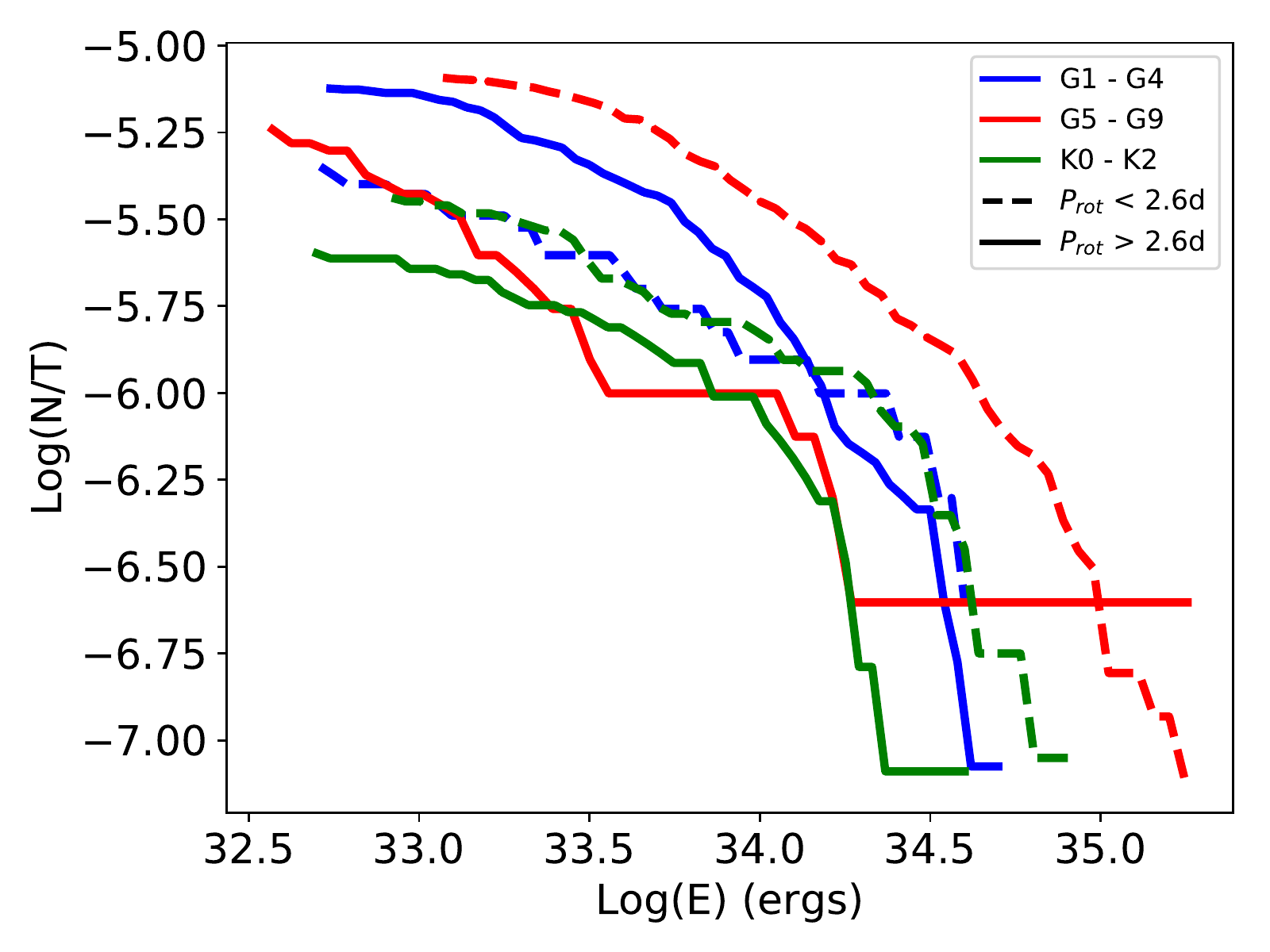}
    \caption{The logarithm of the cumulative flare frequency in seconds against the logarithm of flare energies for a handful of solar-type stars in our sample. These stars have varying spectral types in the ranges G1 - G4 (blue), G5 - G9 (red) and K0 - K2 (green), where each line represents a single star. In addition, these stars have different rotation periods identified by the line style where dashed represents fast rotating with $P_{rot}$ < 2.6 days and solid as slowly rotating with $P_{rot}$ > 2.6 days. }
    \label{freq_energy}
\end{figure}

As well as calculating the flare energies, we want to investigate the frequency of the superflares. Figure \ref{freq_energy} shows the cumulative flare frequency distribution (FFD) of a handful of stars ranging in rotation period and spectral type. The dashed line represents stars with rotation periods less than 2.6 days and a solid line greater than 2.6 days, where 2.6 days represents the median. In addition, stars with varying spectral types are displayed by colour with G1 -- G4 blue, G5 -- G9 red and K0 -- K2 green, where each line represents an individual star. These stars were selected as they possessed a reasonable number of flares for each of the spectral sub-types and had varying rotational periods. Now we will go into more detail about the various stars and the relationship between all six as a whole. 

Firstly, as an example, the dashed red line represents the G6 type star HD 39150 (TIC 364588501) which has a rotation period of 2.28 days and 207 flares. HD 39150 will produce a flare of energy $10^{35}$~erg approximately every 49 days whereas a flare of $10^{33}$~erg will be produced every 1.5 days. Overall, the higher the flare energy the less frequently it will be observed from the star which is consistent with the mechanism known to generate solar flares on the Sun. Prior to the flare energy release the magnetic field becomes stressed and twisted allowing for the buildup of magnetic energy. The flare is then released as thermal energy, kinetic energy and particle acceleration when the magnetic field reconfigures and simplifies through magnetic reconnection \citep[and references therein]{fletcher2011observational}. Therefore, it should take longer to build up and store the magnetic energy required for larger energy flares of $10^{35}$~erg and greater, although see Section 4.5 on flare waiting times.

Looking at the slowly rotating stars as a whole, the G9 type star HD 31026 (TIC 077371445; solid red line in Figure \ref{freq_energy}) which has a $P_{rot}$ = 4.8 days and a total of 23 flares, shows a higher flare energy than the other slowly rotating stars. Oddly enough, this star was only observed in two sectors in comparison to the other slowly rotating stars at six sectors. From the FFD we can see it will take approximately 150 days for this star to produce another of the highest energy flares which is approximately 5 \tess\ sectors of observations. Therefore, it could be coincidence that \tess\ was observing this star at the right time to observe such a large flare. 

In terms of the fast rotating stars, HD 39150 (TIC 364588501; mentioned previously) has the highest flare rate and highest energy, however, it was observed in all 13 sectors. The remaining fast rotators both have the same distribution where the difference in spectral type has no effect. As a whole, the behaviour between the fast and slow rotators does change in each of the spectral type groups, where fast rotators show flares more frequently. However, within the G1 -- G4 spectral group the flare frequency at lower energies is higher in the slowly rotating star.

Overall, for our small sample the  FFD shows the spectral type of the stars does not affect the flare energies. However, it is apparent the rotation period does play a role with the faster rotating stars producing flares of higher energies and flares more frequently. This is to be expected as faster rotating stars tend to produce high energy and overall flare more frequently (with exception of the UFRs) in comparison to their slowly rotating counterparts \citep{hartmann1987rotation, maggio1987einstein}.

\subsection{Flare Effects on Habitability}
The Sun can produce solar flares with energy outputs of up to $10^{32}$~erg. These explosive events have effects which sweep through the entire solar system. Due to Mercury's close proximity with the Sun, its very thin exosphere is constantly being stripped away and then replenished by particles from solar flares, Coronal Mass Ejections and the solar wind \citep[][and references therein]{guinan2004evolution}. On Earth, the magnetosphere protects the planet from harmful radiation and fast moving particles which originate from the Sun. In the event of a large X-class solar flare satellites can be disrupted and aurora can be seen at lower latitudes. Gas giant planets such as Jupiter and Saturn have strong internal magnetic fields and large magnetospheres so the effects of flaring activity can result in aurora and geomagnetic storms \citep{engvold2018sun}. 

Our sample of 209 solar-type stars was cross referenced with the NASA Exoplanet Archive \footnote{\url{https://exoplanetarchive.ipac.caltech.edu}} to identify if any of the targets have known planets. In our sample only one solar-type star, HD 44627 (TIC 260351540), is a host to known exoplanets with a spectral type and rotation period of K1 and 3.9 days. This particular planet is a wide orbiting giant with a semi-major axis of 275AU and mass of 13.5 $M_{JUP}$ \citep{chauvin2005companion}. This star produces flares with energies between $6.1 \times 10^{32}$ - $3.4 \times 10^{35}$~erg flaring 48 times over a period of approximately 160 days. Jupiter is at a distance of approximately 5.5AU from the Sun, therefore, the distance between the star and planet in the HD 44627 system would mean any stellar flares would have a minimal effect. In addition, as the planet is a gas giant and will more than likely possess a considerable magnetosphere similar to Jupiter. 

However, there may be undiscovered exoplanets orbiting other solar-type stars in our sample. Overall, the effects of superflares on orbiting planets will depend on many factors including the star-planet separation, the energy of the flare and the composition of the planet. For example, the flux deposited by a superflare of energy $10^{35}$~erg on a rocky planet at a distance of 1AU will not cause any geophysical alterations. If this planet was icy, however, then it could cause melting which would result in flood plains \citep{schaefer2000superflares}. Another key factor is the potential for the planet to have a magnetic field strong enough to provide protection from harmful flare events. The radiation alone from these events could cause planetary atmospheres to be constantly modified making it difficult for hosting life \citep[and references therein]{vida2019flaring}.

\begin{figure}
    \centering
    \subfloat[]{\includegraphics[width = 0.47\textwidth]{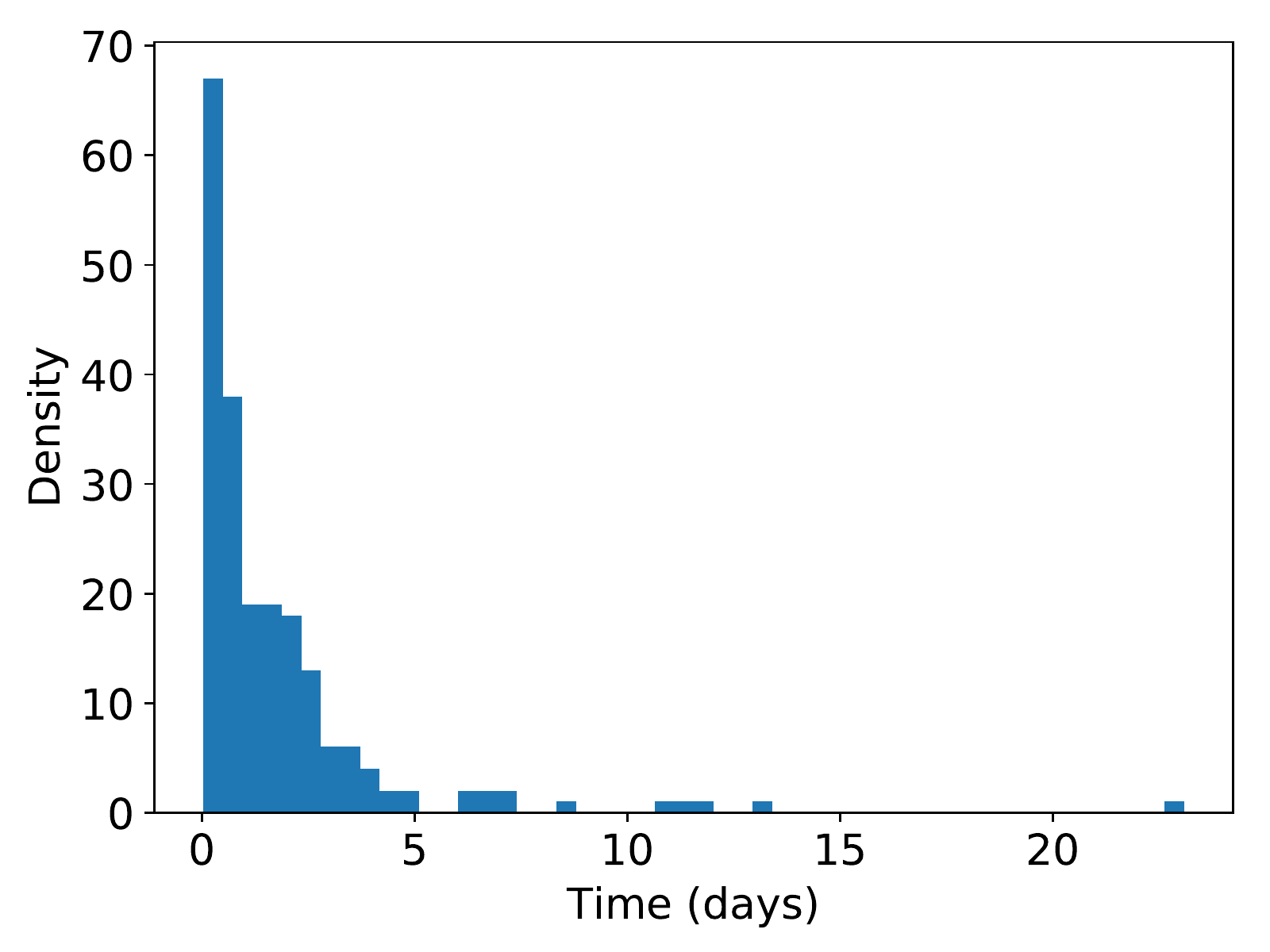}}\\
    \subfloat[]{\includegraphics[width = 0.47\textwidth]{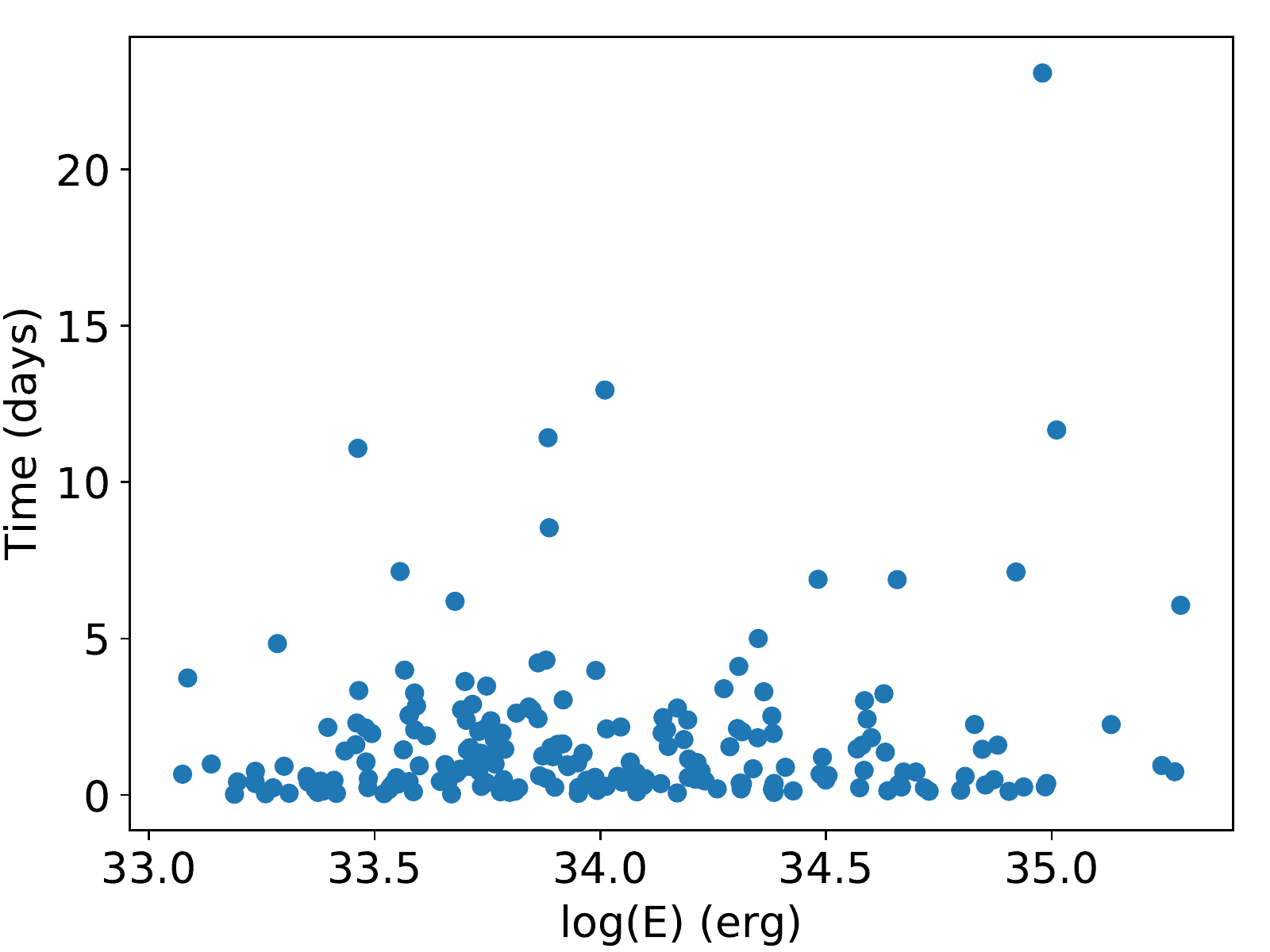}}
    \caption{The distribution of flare waiting times (top panel) for HD 39150 (TIC 364588501) covering 207 flares durig one year of TESS 2-min cadence data. Additionally, the flare waiting times as a function of energy are also shown (bottom panel) detailing no strict correlation between flare magnitude and waiting time. }
    \label{waiting_times}
\end{figure}

\subsection{Flare Waiting Times}
\cite{hudson2020correlation} investigated the flare waiting times between successive solar flares from two active regions AR 10930 (December 2006) and AR 7978 (June 1996). The purpose of this study was to establish the build up/release processes which should be present as a correlation between the waiting time and flare magnitude. Overall, \cite{hudson2020correlation} does observe such a correlation which solidifies the build up/release scenario for solar flares where solar flares result from the build up of magnetic energy in the corona. 

Similar work has been carried out in a stellar flare case, \cite{hawley2014kepler} study the flare waiting times for the low mass M dwarf GJ 1243 using two months of Kepler data. They observe a decline in the number of flares with greater waiting times between the range of 30 minutes -- 8 hours. As a result, they conclude GJ 1243 maintains a steady state of flaring activity which is consistent to a scenario of a number of active regions on the disk. This means some active regions are in the stages of release/decay while others are in the build up phase. 

We conduct a similar analysis using the solar-type star HD 39150 (TIC 364588501) which shows 207 flares within a year of TESS 2-min cadence observations. Overall, a similar decline in waiting times (Figure \ref{waiting_times}(a)) is observed in comparison with \cite{hawley2014kepler}, however, the range in waiting times is much greater on a scale of days. According to \cite{hudson2020correlation}, there should be a correlation with regards to waiting time and flare magnitude (i.e. energy). However, while some flares with longer waiting times do show higher energies, there is a wide spread amongst waiting time and energy as a whole, see Figure \ref{waiting_times}(b). Therefore, this is consistent with HD 39150 possessing multiple active regions on the disk at various stages in the build up/release of flaring activity. We go on to discuss this in more detail in the subsequent sections. 

\section{Long Term Stellar Variability}
\label{stellar_var}
The continuous viewing zone is an area of the sky where each of the \tess\ sectors around the poles in the southern and northern hemisphere overlap. This produces a region of sky with many stars being observed for approximately one year. In our sample of solar-type stars we have two which have been observed in the continuous viewing zone, HD 39150 (TIC 364588501) which was observed in all 13 sectors and HD 47875 (TIC 167344043) which was observed in all 13 sectors minus sector 11 where no data was collected. This provides lightcurves of these stars which span one year allowing us to investigate long-term levels of variability within these stars. 

\begin{figure*}
    \centering
    \includegraphics[width = 1.0\textwidth]{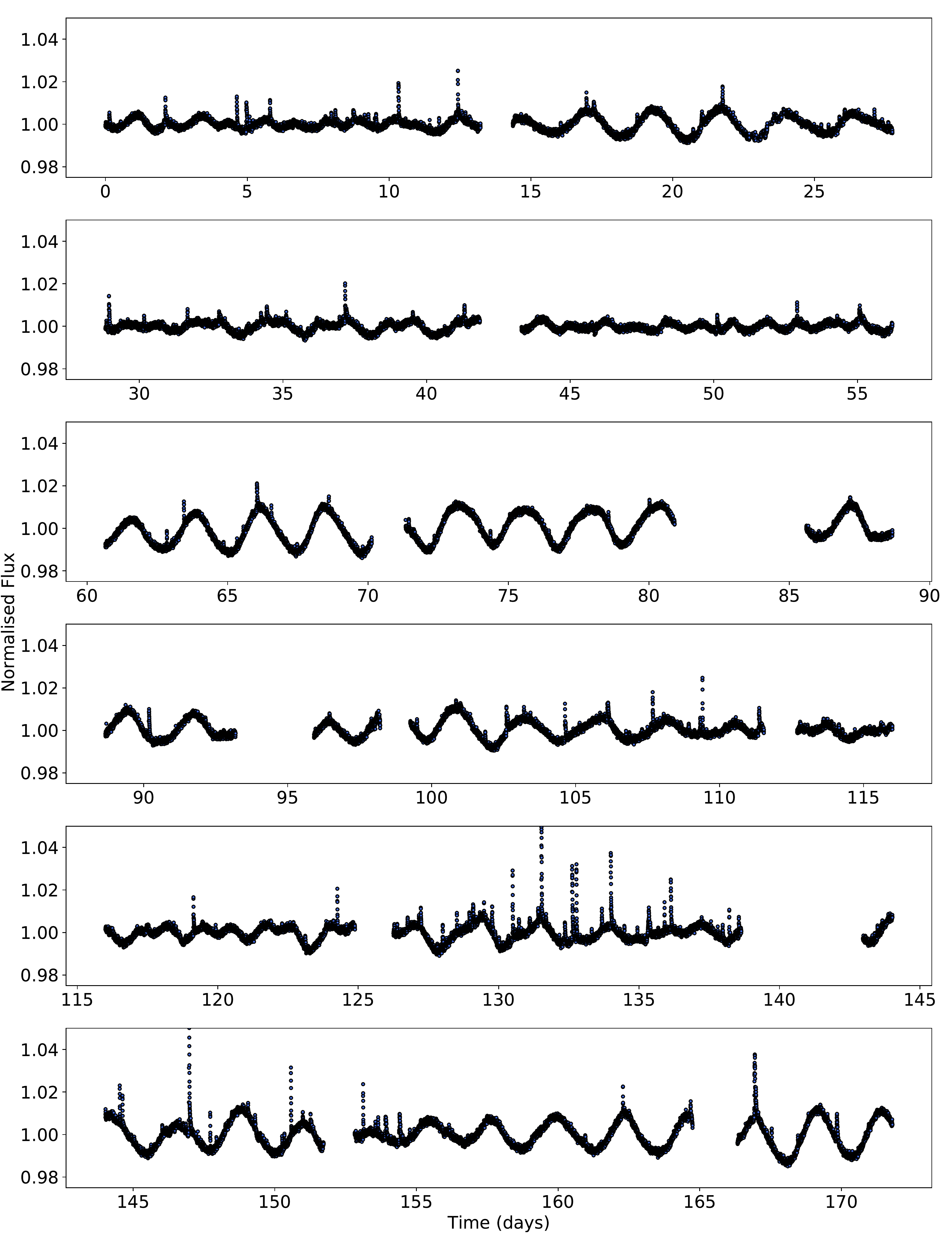}
    \caption{The lightcurve for HD 39150 (TIC 364588501) covering approximately 170 days detailing the magnetic variability of the star including changing spot structures and flaring. }
    \label{var_fig}
\end{figure*}

HD 39150 (TIC 364588501) which was observed for a total of 357 days. This G6 star has a rotation period, $P_{rot}$ = 2.28 days and a total of 207 flares with energies reaching $10^{35}$~erg. Figure \ref{var_fig} shows the lightcurve of this star spanning approximately 140 days, detailing the changing nature of the rotational modulation. In the bottom left and right panels there is evidence of multiple spots as the shape of the modulation changes. This could be the result of two active regions possessing spots which are rotating with marginally different periods, meaning they are slightly unsynchronised. The whole \tess\ lightcurve of this star from all sectors can be folded to one rotation period and phase zero which also suggests this scenario. In addition, the amplitude of the rotational modulation changes sector to sector suggesting the active regions are growing and decaying as the star rotates or new regions are developing/disappearing. To determine if some of the variability was due to instrumental effects we examined lightcurves of spatially nearby stars and found no variation between different sectors. Therefore, we conclude the variability in HD 39150 (TIC 364588501) is intrinsic to the star. With regards to flaring activity, there seems to be an increased level towards the end of this section of lightcurve between days 125 and 155 where the flares appear more frequently and with a greater energy. \cite{tu2019superflares} also discuss this target noting the increase in flaring activity within sector 5, however, they do not offer any explanation regarding the reasoning behind this. We discuss this star further later in Section \ref{phase_individual} providing a potential explanation for the sudden increase in activity while also associating it with the rotational phase distribution of the flares. 

HD 47875 (TIC 167344043) is also present within the continuous viewing zone and was observed for a total of 330 days. This star has a spectral type of G4, $P_{rot}$ of 2.99 days and a total of 179 flares with energies in the $10^{34}$~erg range. The rotational modulation of this star is constant throughout the year of observations possessing a clear sinusoidal pattern with no evidence of multiple spots. However, the amplitude of this particular star also changes producing a multiperiodic lightcurve. This would suggest there are potentially migrating spots on the disk of the star which fall into differential rotation. As a result of this phenomena, large flares are observed as the migrating spot crosses the disk of the star.

Overall, there appears to be more variability observed within solar-type stars on timescales of months, with regards to their spotted structures, as compared to low mass stars. We know the active regions and spot structure observed on the Sun change over periods of weeks to months with no sunspots lasting years. Therefore, it is not unexpected to observe this behaviour in other stars of a similar spectral type. As \tess\ returns to the southern ecliptic in Cycle 3, follow ups of these stars would be valuable to continue to monitor the changing behaviour observed. This will then lead into long term observations with the potential to determine stellar cycles which is important in understanding the overall magnetic cycle on other stars. 

\section{Starspot Areas}
Determining the areas of starspots is a non-trivial process and there are many ways to do so including Zeeman Doppler Imaging \citep{rosen2015first}, Spectral Modelling \citep{fang2016stellar, gully2017placing} and Planet-Transit Spot Modelling \citep{morris2017starspots}. In addition, it is possible to use the amplitude of the rotational modulation from lightcurves to provide a rough indication of the approximate areas of starspots on the stellar disk \citep{rebull2016rotation,rebull2016rotation2, giles2017kepler}. However, this process underestimates for the presence of polar spots, circumpolar spots, bands of spots, spots all over the disk and a pole on star with spot distributions. Despite this, there is still merit in determining starspot areas as it can provide some insight into the conditions needed for these large energy superflares.

\begin{figure}
    \centering
    \subfloat{\includegraphics[width = 0.47\textwidth]{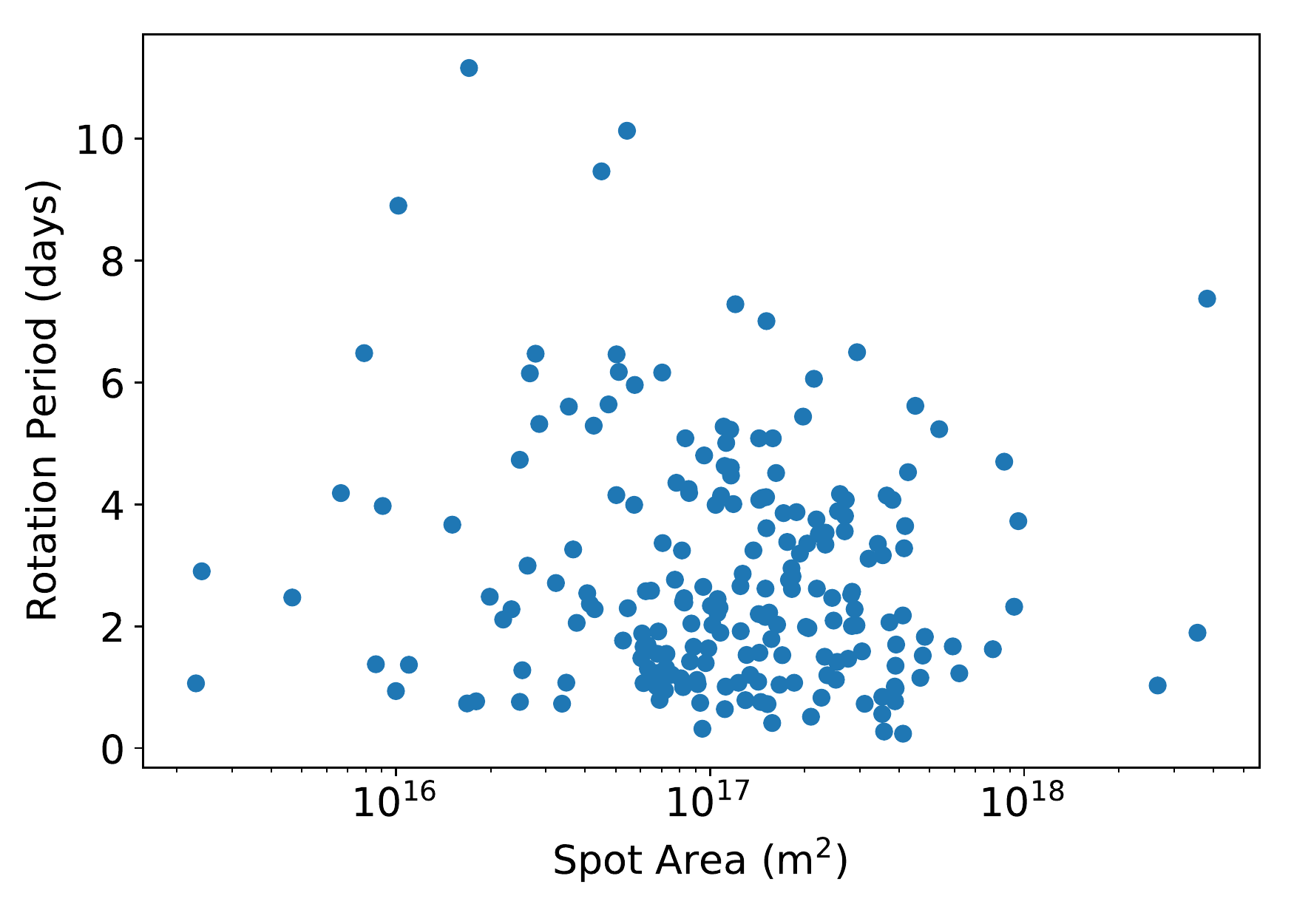}}\\
    \subfloat{\includegraphics[width = 0.47\textwidth]{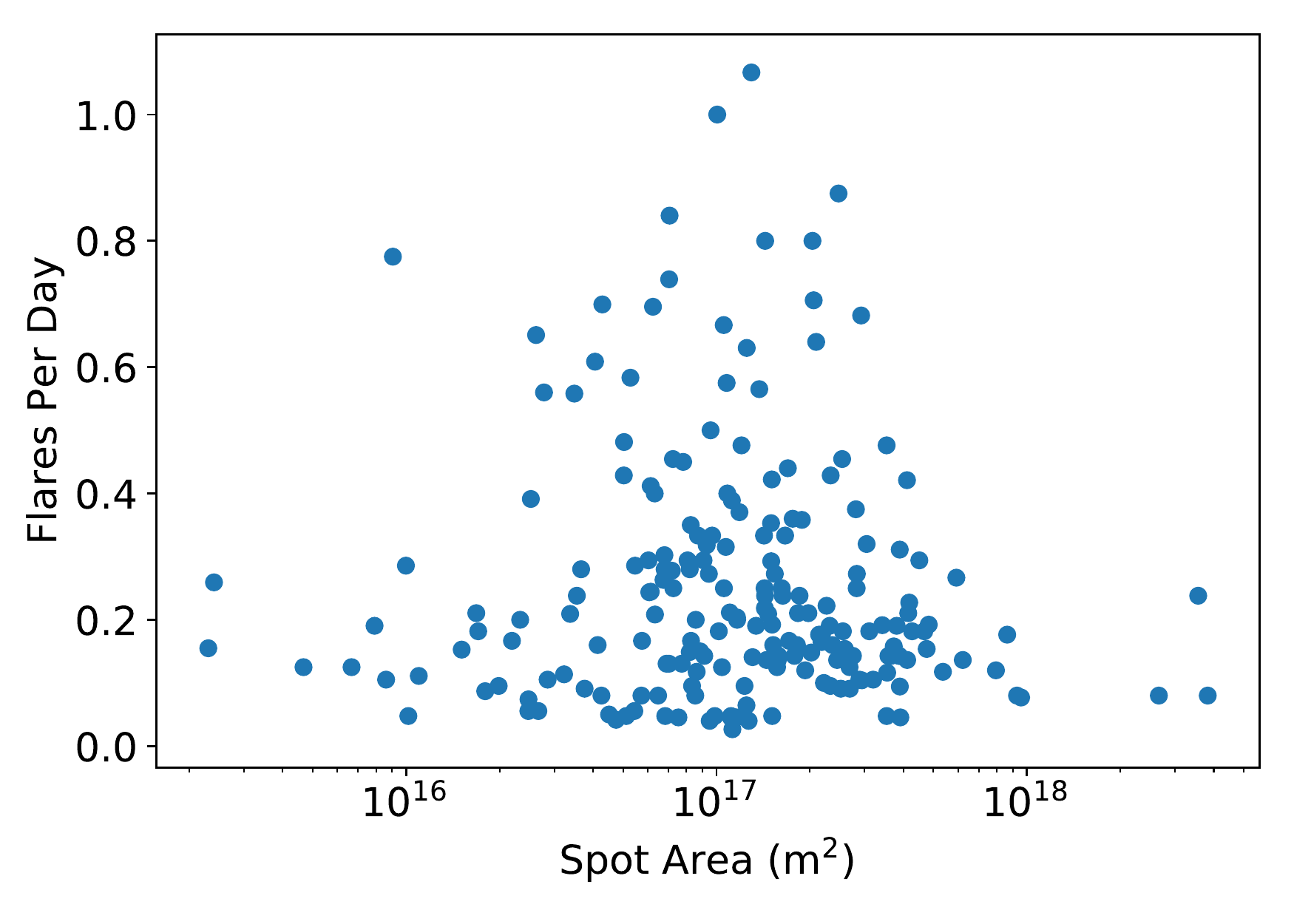}}\\
    \subfloat{\includegraphics[width = 0.47\textwidth]{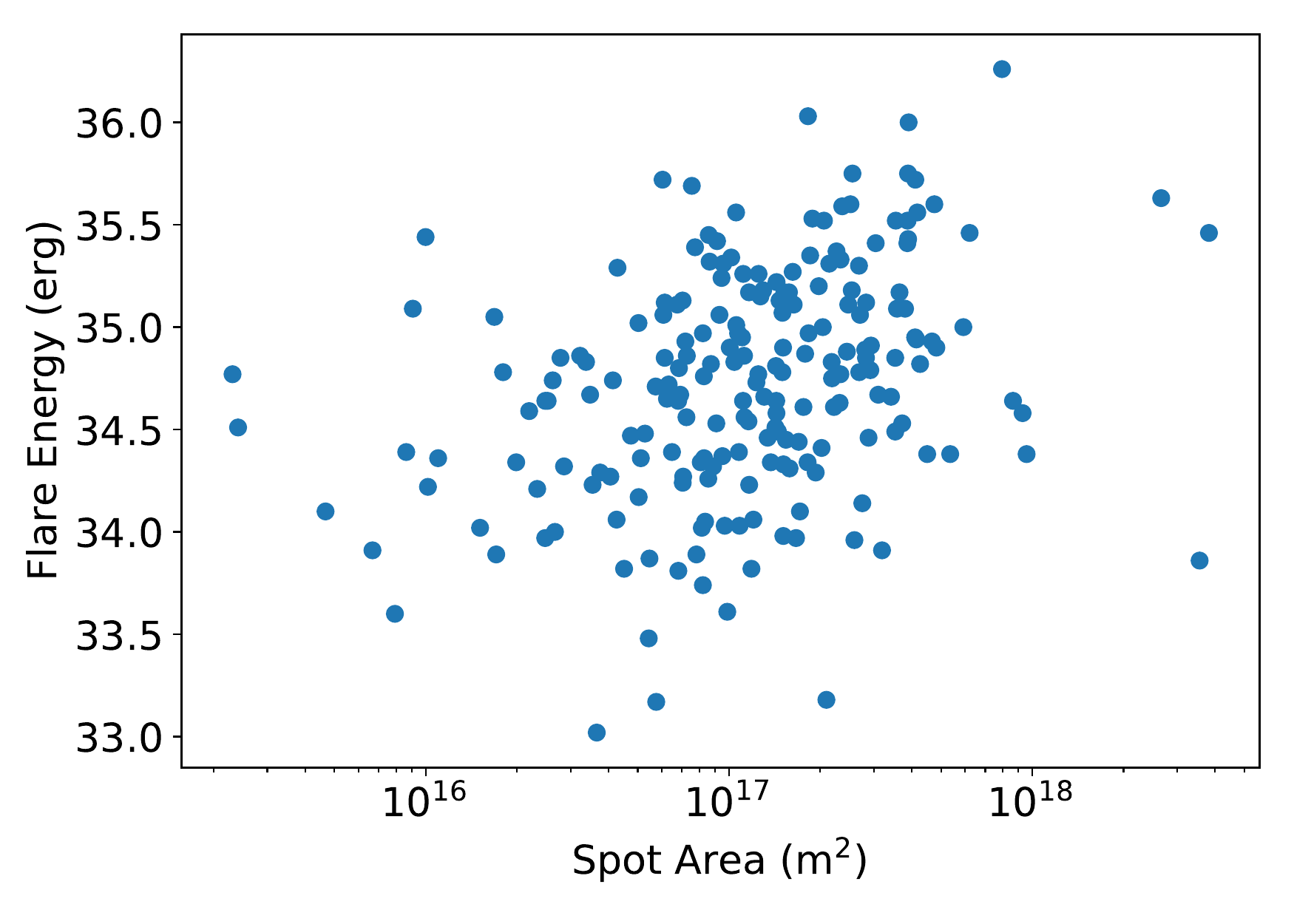}}
    \caption{Here we show the spot coverage of each star calculated from the amplitude modulation of the \tess\ lightcurves as a function of stellar rotation period (top panel), normalised flares per day (middle panel) and maximum flare energy (bottom panel). }
    \label{starspot_area}
\end{figure}

In order to determine the starspot area we will look to use the method described by \cite{notsu2019kepler}. Firstly we must estimate the temperature of the spot which can be done by applying a relation on the difference between the photosphere and the spot. The relationship is as follows:

\begin{equation}
    T_{star} - T_{spot} = 3.58 \times 10^{-5}T_{star}^2 + 0.249T_{star} - 808
\end{equation}
\vspace{1mm}

where $T_{star}$ is the effective photospheric temperature obtained from Gaia DR2 and $T_{spot}$ is the temperature of the spot. Next, we use $T_{spot}$ to calculate the area of the starspot according to the variations within the lightcurve as follows:

\begin{equation}
    A_{spot} = \frac{\Delta F}{F}A_{star} \bigg[1 - \bigg(\frac{T_{spot}}{T_{star}}\bigg)^4\bigg]^{-1}
\end{equation}
\vspace{1mm}

where $\Delta F/F$ is the amplitude of the normalised lightcurve, which was measured from the phase folded and binned lightcurve and $A_{star}$ is the area of the stellar disk calculated as $2\pi R^2$ ($R$ is taken as the Gaia DR2 radius). As an example and benchmark we calculated the starspot temperature for the Sun, a G2 type solar star with a temperature of $T_{star}$ = 5800K, finding $T_{spot}$ = 3960K. This temperature is reasonable and aligns with the temperature range for sunspots at 3500K - 4550K \citep{solanki2003sunspots}. Using these relationships, we are able to determine approximate spot areas for all 209 solar-type stars within our sample.

According to \cite{mcintosh1990classification}, larger more energetic solar flares occur from active regions which host larger spot coverages and complex spot structures. Therefore, we would expect to see the same behaviour within other solar-type stars. In Figure \ref{starspot_area}, we look at the stellar spot coverage as a function of rotation period, flare number and flare energy to investigate whether the predicted behaviour is in fact observed. With regards to rotation period, you would expect a larger spot coverage to result from a faster rotation period. However, in our sample there appears to be no relation and overall there is a large spread in both rotation and spot coverage. This was tested further using the Pearson correlation coefficient (P) which yielded a result of P = --0.13 also indicating the lack of any correlation. The Pearson correlation coefficient is a statistical test which measures the linear relationship between two variables. A coefficient of P=+1 indicates a direct positive linear correlation and P=--1 a direct negative linear correlation, with  P=0 indicating no correlation.

Next we look at the number of flares as a function of spot area. The expectation is that the larger the spot coverage the more flares will be observed from the star. However, this is not the case and we observe a peak in flare number at approximately $10^{17}$ m$^2$ where afterwards, there appears to be a drop in flare number for the larger spot coverages. Could this be because these larger spot coverages are producing higher energy flares less frequently? Again, this was tested with the Pearson correlation coefficient providing a value of P = -0.05 which indicates there is no linear relationship present. This then brings us on to look at the energy of the flares as a function of spot coverage. Although there appears to be some evidence for higher energy flares resulting from larger spot coverages (Figure \ref{starspot_area}), the Pearson correlation coefficient provided a result of P = 0.38 indicating there is a slight significant linear correlation present. Overall, none of the Pearson correlations for any of the plots in Figure \ref{starspot_area} indicate a strong linear correlation.

The solar-type star CPD-5711 31 (TIC 279614617) is a G8 type star which was observed to have the largest starspot coverage despite only having a rotation period of 7.37 days. It was observed in \tess\ sector 1 and produces two flares during this time. This is a relatively low number considering it has the largest spot coverage of $3.8 \times 10^{18}$ m$^2$ (which equates to 10\% of the visible stellar disk) from the solar-type star sample. However, it does produce two of the larger flares with energies in the $10^{35}$~erg range. 

A similar analysis was conducted by \cite{howard2019evryflare} and \cite{notsu2019kepler} where they also investigated the spot coverages of their stellar samples against various flare properties. In \cite{howard2019evryflare} they use photometric data from Evryscope \citep{law2015evryscope} lightcurves of 113 cool stars to investigate rotation periods, starspot amplitudes and flare properties. They did not find a relationship between the size of the spot coverage and the energy of the flares produced but were able to constrain the minimum field strength of their late K to mid M flare stars as 0.5kG. In \cite{notsu2019kepler} they conduct an investigation into the relationship between superflares and rotation period, including an analysis on the spot coverages of their sample of solar-type (G-type) stars. Overall, they do see a relationship between flare energy and starspot coverage, concluding the superflare energy is in fact related to the starspot coverage of the star. In addition, they also see that superflares tend to occur from stars with shorter rotation periods and larger starspot coverages. 

\section{Rotational Phase}
One of the criteria for our sample was the presence of rotational modulation within the \tess\ lightcurve. This rotational modulation is the result of starspots which are present on the stellar disk and move in and out of view as the star rotates. In Papers 1 and 2 we test the distribution of flares in samples of M dwarfs using a simple statistical test. Our findings show no evidence for any preference in rotational phase, indicating the flares are randomly distributed. In solar physics there is a well-established relationship between sunspots and solar flares. Therefore, it is surprising to find no such correlation amongst other flare stars. 

In this study, we will use the same simple $\chi_{\nu}^{2}$ test to assess the phase distribution of the flares. We will look at the flares from all 209 solar-type stars as well as several stars which possess the highest flare rate. In addition, stars which possess evidence of multiple spots within their lightcurve will be extracted and a separate analysis will be carried out on the remaining stars. All of this will allow a thorough analysis on the rotational phase of the flares, determining whether in solar-type stars there is a starspot/flare relationship similar to the Sun.

\subsection{The Overall Flare Sample}
\label{sec:phase_all}
\begin{figure}
    \centering
    \includegraphics[width = 0.47\textwidth]{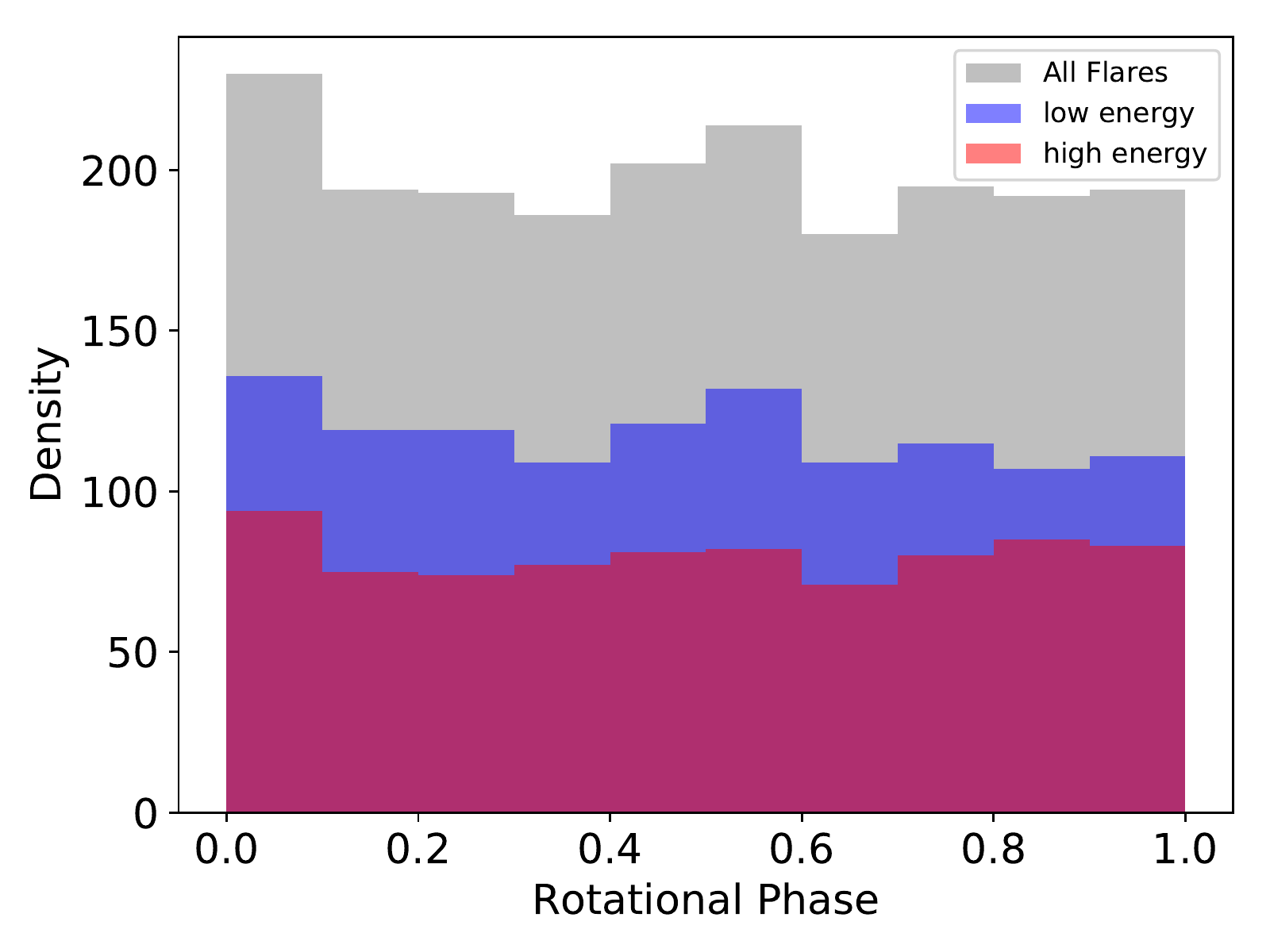}
    \caption{A histogram of the rotational phase distribution for all 1980 flares from the sample of 209 solar-type stars. The bin size is $\phi = 0.1$ and the spread shows no preference for any rotational phase. The cut-off for low and high energy flares was determined as $10^{34}$~erg.}
    \label{phase_all}
\end{figure}

Taking all 209 solar-type stars in our sample with a total flare number of 1980, we can test for any preference in rotational phase. For all of our stars phase zero, $\phi = 0.0$, is defined as flux minimum of the rotational modulation allowing for this comparison. Utilising a simple $\chi_{\nu}^{2}$ test (see Paper I for full details), no correlations between flare number and rotational phase was found. The values of the $\chi_{\nu}^{2}$ test are 1.22, 1.04 and 1.04 for all, low and high energy respectively, where the cut-off was determined as $10^{34}$~erg according to the distribution of flare energies in Figure \ref{hist_flares}. This indicates the flares are randomly distributed and do not coincide when the starspot is most visible. Figure \ref{phase_all} shows the histogram distributions for the flares with phase bins of $\phi = 0.1$, where a consistent spread of flares is present amongst all, high and low energies. In addition, Kolmogorov-Smirnov (KS) and Shapiro-Wilk (SW) tests were also conducted with KS $= 0.5$ with $p-value = 0$ and SW $= 0.95$ and $p-value = 1.49 \times 10^{-24}$. The results of these tests again show the flares are randomly distributed and are in agreement with the $\chi_{\nu}^2$ test. 

In \cite{roettenbacher2018connection} they conduct a similar analysis for a group of 119 main sequence stars from late-F to mid-M. However, they do find a correlation between rotational phase and flare number which is presented as a peak in their histogram plot similar to Figure \ref{phase_all}. This correlation is only present in flares which have flux increases between 1\% and 5\% and so is not present in higher energy flares. Their sample consists of mid-M type stars, while ours does not, furthermore they use \kep\ long cadence (30-min) observations whereas we have \tess\ 2-min cadence. These differences along with the addition of omitting stars which show any evidence for multiple spot structures in their analysis could be why our results differ. 

As mentioned previously there is a large variation in the rotational modulation of the \tess\ lightcurves within the solar-type sample caused by the presence of multiple spots. In our sample $\sim$ 40\% show a clear sinusoidal pattern as a result of one starspot structure present on the disk of the star. The remaining 60\% show evidence of multiple spot structures which could potentially cause controversy within the rotational phase findings of all flares from all stars. To address this we remove all stars which show any potential evidence for multiple spot structures and conduct the test on the remaining sample. This consists of 83 solar-type stars with a total of 886 flares. Our results, again, show no preference for rotational phase both in the $\chi_{\nu}^{2}$ test, KS test and SW test. Therefore, this strengthens our conclusion that the flares do not originate from the dominant spot/active region but are randomly distribution in rotational phase. 

\subsection{Individual Case Studies}
\label{phase_individual}
We have selected two solar-type stars from our sample which have the highest flare rate and therefore, are ideal candidates to investigate the relationship between rotational phase and flare number. Both of these stars are present in the continuous viewing zone of the \tess\ mission and their stellar variability was discussed in Section \ref{stellar_var}. However, in this section we will only be focusing on the distribution of flares within their lightcurves as a function of rotational phase. These stars are HD 47875 (TIC 167344043) and HD 39150 (TIC 364588501) with spectral types G4 and G6, flare numbers of 179 and 207 and rotation periods, $P_{rot}$, of 2.99 and 2.28 days, respectively. We utilised the same simple $\chi_{\nu}^2$ test as discussed previously obtaining values for all low and high as of 1.55, 1.41 and 1.11 for HD 47875 and 0.75, 0.89 and 0.76 for HD 39150. As a result, we find no significant evidence (>$3\sigma$ confidence) of any correlations between rotational phase and flare number.

\begin{figure*}
    \centering
    \subfloat[]{\includegraphics[width = 0.47\textwidth]{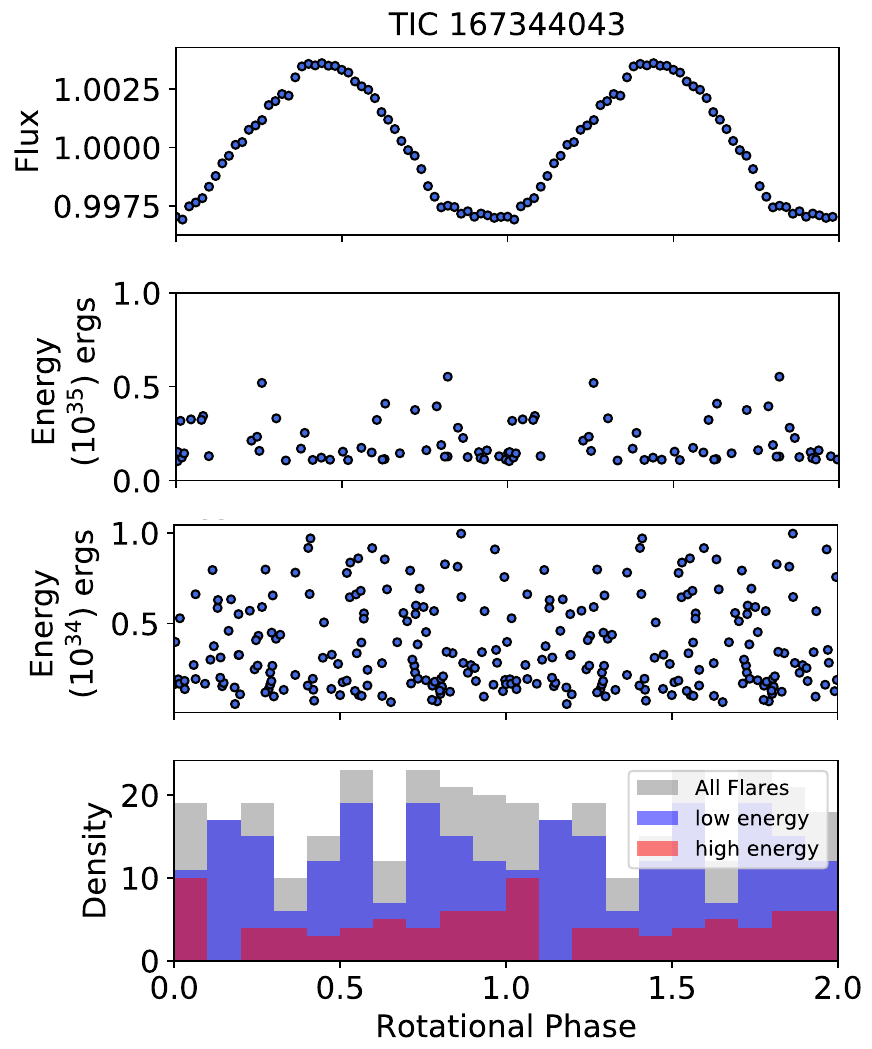}}
    \subfloat[]{\includegraphics[width = 0.47\textwidth]{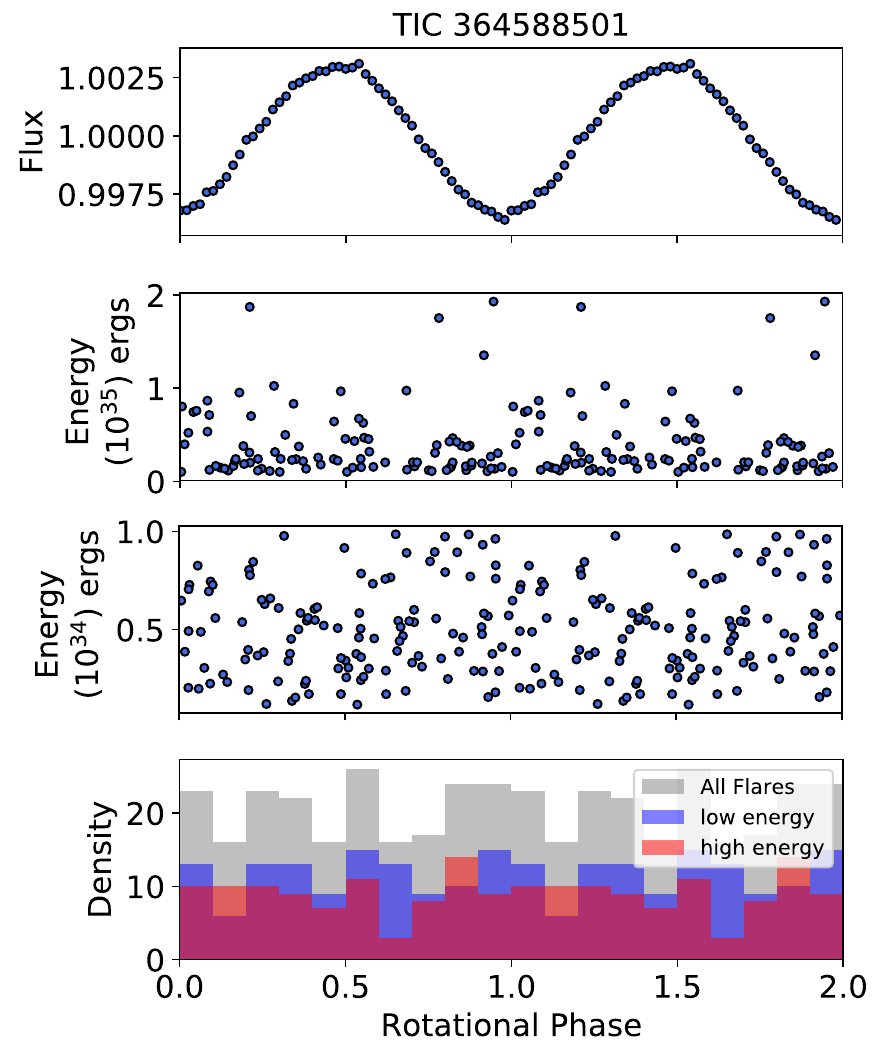}}
    \caption{Here we show phase folded and binned lightcurves (upper panel) along with the rotational phase distribution of the flares vs. their energy (middle panels). The top energy plot represents higher energy flares with $E_{flare}$ > $10^{34}$~erg and the bottom $E_{flare}$ < $10^{34}$~erg. In addition, we also show a histogram distribution of the flares as a function of rotational phase for completeness. The rotational phase coverage $\phi = 1.0 - 2.0$ is simply a repeat of $\phi = 0.0 - 1.0$. (a) HD 47875 (TIC 167344043) is a G4 star with $P_{rot} = 2.99$ days and a total number of 179 flares observed over 12 \tess\ sectors. (b) HD 39150 (TIC 364588501) is a G6 star with $P_{rot} = 2.28$ days and 207 flares observed over 13 \tess\ sectors.}
    \label{individual_phase}
\end{figure*}

In Figure \ref{individual_phase} we show the distribution of the flares as a function of their energy along with the phase folded and binned lightcurves. From these plots, it is easier to see there are flares present at all rotational phases in both high and low energy. In addition, the flares are randomly distributed and there is no preference for any rotational phase even during rotational minimum when the dominant starspot is most visible. Although, there is a hint of increasing high energy (> $10^{34}$~erg) flares at phase 0.25 and 0.75. 

Due to the length of the \tess\ observations for HD 39150 we want to approach the analysis of flares and rotational phase in a different manner than our previous studies. From Figure \ref{var_fig} we see that within a few days (e.g. from day 137 to day 140), a large change in the spot structure appears. At day $\sim$145, we have the presence of a large dominant spot structure while a few days earlier we have the emergence of multiple spots and plage activity which reduces the spot contrast. However, the most interesting aspect is the intense flare activity from day 127 to day 138 which clearly indicates a strong link between emerging spots and flare activity. After this point the star returns to a less active state similar to what is observed at the beginning of the lightcurve. \citet{chandra2010flaring} showed that the triggering mechanism for intense flare activity was a combination of flux emergence, shearing between the magnetic polarities of the two flux systems (emerging and pre-existing) plus the interaction of the new emerging bi-poles with pre-existing field. With the spot structure for HD 39150 (TIC 364588501) changing rapidly on a daily basis, intense flare activity is expected.

As a result of this we computed the $\chi_{\nu}^2$ test on the sectors of \tess\ data for HD 39150 on an individual basis and focused on the period in sectors 5 and 6 which showed the intense flaring activity. However, the values for the $\chi_{\nu}^2$ test show no preference for rotational phase. This is to be expected as by looking at the lightcurve in Figure \ref{var_fig} during this active period there are flares present at all rotational phases. Despite this, the flares are present as a result of the emergence of multiple spots and plage activity which does suggest the link between spots and flares. However, the spot structures on these stars are more complex than a simple one spot model and so this would be the main reason behind a lack in any clear correlation. 

\section{The Solar Analogue}
With the Sun being our nearest star, astronomers and physicists have been collecting detailed spatial observations to study its phenomena for nearly 150 years. Similarly, there is a wealth of historic data including sunspot number and flare properties dating back to the 1930s. Overall, these data sources provide a deep knowledge of our closest star aiding in the understanding of multiple solar phenomena and its effects on the Earth and Solar System. In this section we utilise historic X-ray data of solar flares from the Geostationary Operational Environmental Satellite (GOES) archive and sunspot numbers from the Sunspot Index and Long-term Solar Observations (SILSO) database to detail the close relationship between solar flares and sunspots.  

\begin{table}
    \centering
    \begin{tabular}{cc}
    \hline
         Flare Classification & Energy Range (ergs)    \\
    \hline
         X10                  & > $10^{32}$            \\
         X                    & $10^{31}$ - $10^{32}$  \\
         M                    & $10^{30}$ - $10^{31}$  \\
         C                    & $10^{29}$ - $10^{30}$  \\
         B                    & $10^{28}$ - $10^{29}$  \\
         A                    & < $10^{28}$            \\
    \hline
    \end{tabular}
    \caption{The energy range for each of the solar flare classifications, making comparisons between the stellar and solar flares easier \protect\cite[obtained from][]{notsu2019kepler}.}. 
    \label{solar_flare_energies}
\end{table}

\begin{figure}
    \centering
    \subfloat[]{\includegraphics[width = 0.47\textwidth]{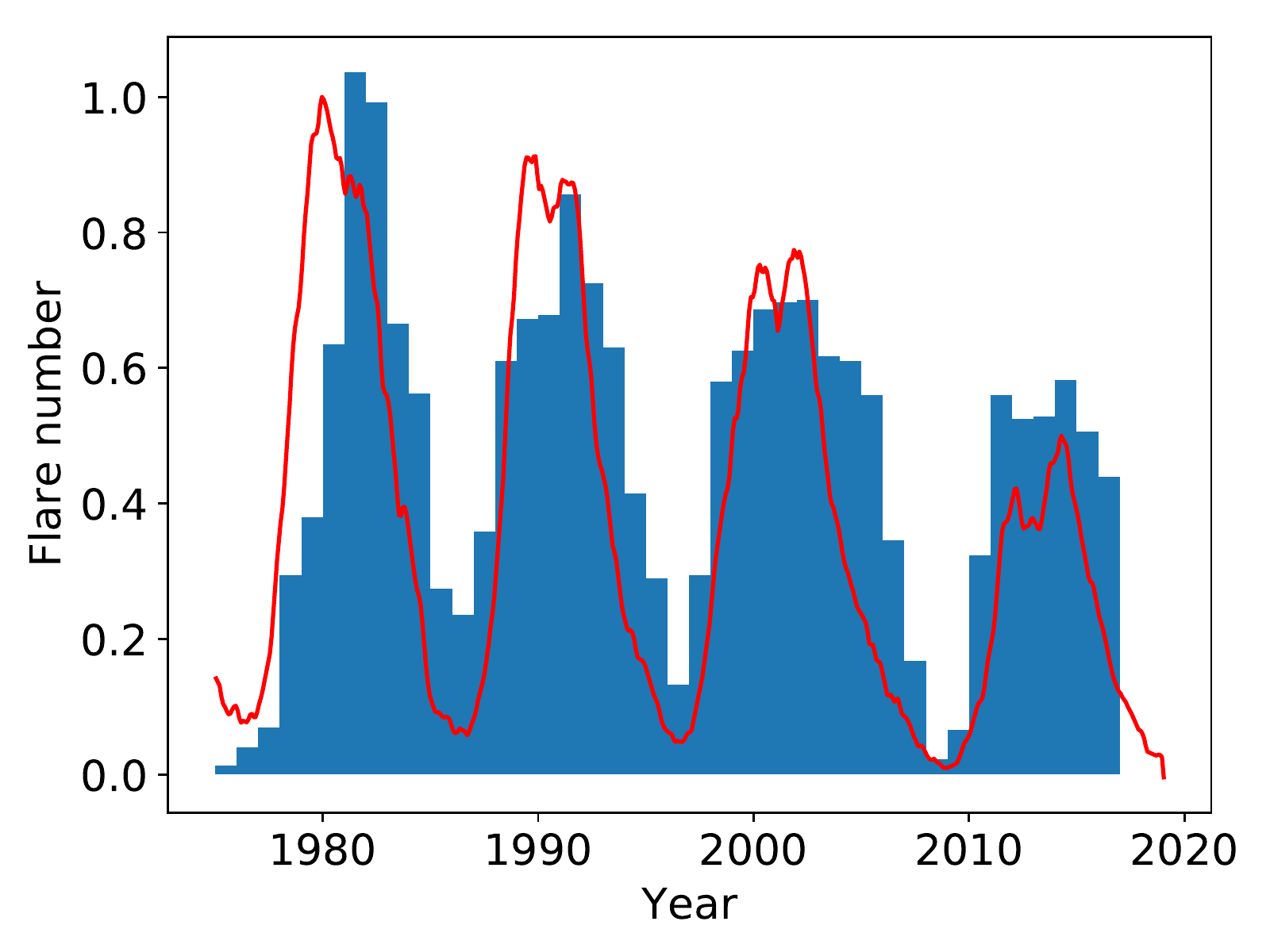}}\\
    \subfloat[]{\includegraphics[width = 0.47\textwidth]{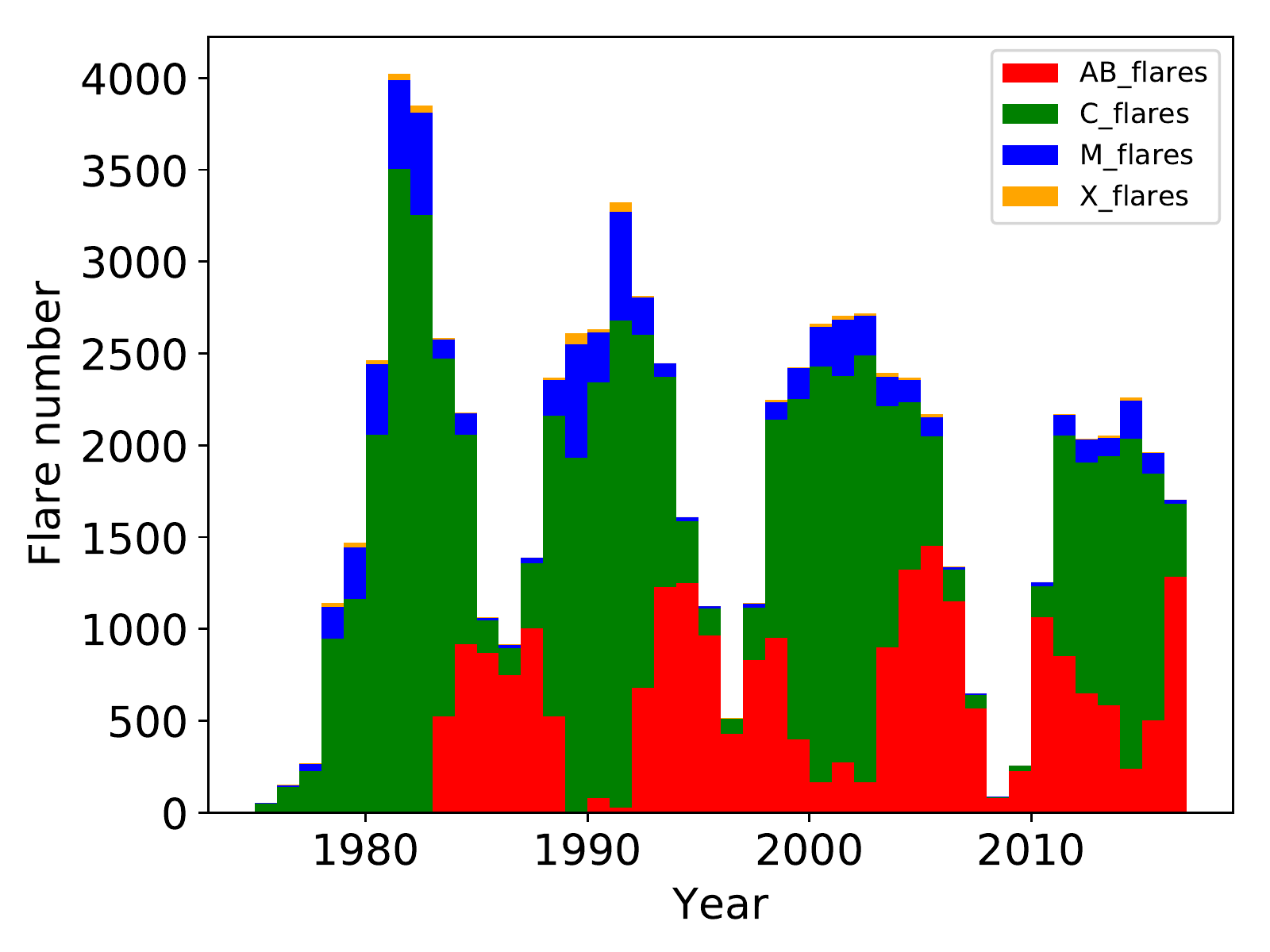}}
    \caption{Here we show two plots detailing the historic data of solar flares and sunspots. Plot (a) shows the sunspot number in the red solid line and the flare number as the blue histogram. These have both been normalised to 1 to allow a comparison of the data revealing the close relationship between sunspots and solar flares. Plot (b) shows the varying sub-classes of solar flares within revealing an anti-correlation between the lower A \& B-class solar flares to the higher energy X, M and C-class solar flares. The data of the solar flares was obtained from the GOES archive and for the sunspots from the SILSO database and all the data has been binned per year.}
    \label{solar_spot}
\end{figure}

The relationship between flares and sunspots is well-established and it is generally accepted these phenomena are closely related. Figure \ref{solar_spot}(a) sums up this close relationship where the sunspot number (red line) and flare number (blue histogram) are observed to be correlated with each other over the solar cycle. However, despite this, the relationship between sunspots and flares is more complicated than initially believed. In \cite{gao2016curious} they investigated the temporal behaviour of varying classes of solar flares. Their findings show the lower B-class solar flares to be in anti-phase with all other C, M and X-class solar flares in terms of the solar cycle (see Table \ref{solar_flare_energies} for details of the various solar class flares and their respective energies). To investigate this strange behaviour, we plot the solar sub-classes as a histogram in Figure \ref{solar_spot}(b) which also displays an anti-correlation amongst the A \& B-class solar flares. Even more interestingly, the A \& B-class flares are also out of synchronisation with the sunspot number. These lower energy (<$10^{29}$~erg) solar flares are present when the Sun is in a solar minimum and is considered not very active where there is very little spot coverage. Therefore, these flares could be originating from plage regions or areas where local dynamos are at play. The higher energy (>$10^{29}$~erg) solar flares are then clearly correlated with sunspots and both appear during solar maximum when the Sun is at its most active. As suggested by \cite{gao2016curious} this anti-correlation within low class flares could potentially be linked to the negative correlation between small and large sunspots \citep{nagovitsyn2012possible}.

Applying these scenarios to our solar-type star sample could aid in explaining the lack of a correlation between rotational phase and flare number. Lower energy flares are present at all rotational phases (see Figure \ref{individual_phase}) which could result from plage, filaments or local dynamo regions not associated with spots which are present across the stellar disk. This is similar to what is observed on the Sun where the lower energy flares are predominately present when the Sun is less active and producing a lower number of spotted regions. The lower energy A \& B class solar flares are a few orders of magnitude less energetic than those observed on the solar-type sample. However, it is possible to observe flares of energies $10^{29}$~erg in solar-type stars, for example using \kep, see \cite{maehara2012superflares}. Therefore, these lower energy flares could be present but not observable with \tess\ since it is less sensitive to lower energy flares (see Section \ref{tess_lightcurves}). As a result we note our flare sample does not contain any flares with energies less than $10^{30}$~erg which equates to a C-class solar flare (see Table \ref{solar_flare_energies}), however, in our solar-type sample a lower energy flare is considered to be <$10^{34}$~erg according to Figure \ref{hist_flares}.

Overall, the lower energy (<$10^{34}$~erg) stellar flares from the solar-type sample could originate from plage, filaments or local dynamo regions on the stellar disk not associated with starspots. This is similar to the lower energy A \& B class solar flares which are also not associated with sunspots. Therefore, the stellar flares observed at all rotational phases could be resulting from a mixture of spotted and non spotted regions resulting in no correlation between rotational phase and flare number. Alongside this, the larger energy flares are more present when there is spot activity present on the disk. This is particularly clear in the star HD 39150 in Figure \ref{var_fig} where emerging spot and plage regions cause increased flaring activity as a result of shearing between varying magnetic polarities. 

All of the discussions so far centre around the relationship between solar flares and sunspots on a cycle basis. The \tess\ observations are far too short to investigate cycle periods on our sample, even the ones observed in all sectors. It will take a few years for \tess\ to build up longer term observations of stars as it returns to sectors. Therefore, can we correlate the solar flare numbers within an activity cycle to the solar rotation period? The above work regarding solar flares discusses X-ray data while with \tess\ we deal with flare observations in the optical. What is required for the Sun is a discussion on whether there is a relationship between solar flares and sunspots on the rotational period basis. Unfortunately, catalogues of solar White Light Flares (WLFs) are very much incomplete: for example \cite{matthews2003catalogue} lists 28 flares over a one year period detected with Yohkoh, a Japanese Solar mission. The catalogue from Kuhar et al (2016) contains 43 M and X class flares which occurred from 2011 to 2015 and were observed by both SDO/HMI and RHESSI and \cite{namekata2017statistical} expanded on this adding another 11 observed in 2015. Overall, this data is not sufficient for a proper analysis of the above question. Moreover, obtaining the rotation period of the Sun as an effective lightcurve in time is a difficult process and despite irradiance measurements of the solar disk being taken this is not something which is well studied in solar physics.

\section{Discussion}
\label{discuss}
In Papers I \& II, we find no evidence for a correlation between rotational phase and flare number within a sample of 183 M dwarf flare stars. We now have a similar study for a sample of 209 solar-type (F7 - K2) stars observed in 2-min cadence by \tess\ in sectors 1 - 13. We determined rotation periods using an LS periodogram, identified and characterised 1980 flares within our sample and determined their energies between $10^{31}$ - $10^{36}$~erg. Similar to Papers I \& II, we find no evidence of any correlation between rotational phase and flare number, indicating the flares are randomly distributed and do not occur alongside the dominant starspot which is responsible for the rotational modulation. However, it was noted that increased levels of activity were observed in the star HD 39150 (TIC 364588501) when evidence of emerging starspot and plage regions were observed in the rotational modulation of the lightcurve. This finding does then suggest there is a relationship between flares and starspots and agrees with the mechanism for flare generation as discussed by \cite{chandra2010flaring}.

In Papers I \& II, we identify four possible scenarios to explain the lack of a correlation including star-planet interactions, binarity, polar spots and multiple spot locations. Here we discuss these further while also bringing in other possible explanations. Firstly, there is the potential for star-planet interactions (SPI's) and interactions between two stars within a binary system. Only one of the solar-type stars in our sample, HD 44627 (TIC 260351540), has a known exoplanet. In this instance the planet is orbiting at a distance (a = 275AU) which is unlikely to cause any SPI's with its host star. However, this does not rule out any of the other stars in our sample having undiscovered exoplanets which could cause SPI's. 

Additionally, the stars in our sample could be in binary systems which could  cause induced magnetic activity producing flares. Within our sample there are three stars whose \tess\ lightcurves show they are eclipsing binaries. These solar-type stars are CD-78 516, AF Cru and HD 120395 (TIC 357911163, $P_{orb}$=1.63 d; TIC 309528896, $P_{orb}$=1.89 d; and TIC 243662768 $P_{orb}$=1.64 d respectively), which have rotation periods between 7.5 -- 10 days, a spread in spectral types and show little flaring activity. The lack of flaring activity in these stars indicates that the period is long enough that interactions between the stars do not give rise to increased flaring activity.

Secondly, with our M dwarf samples in Papers I \& II we discuss the potential for polar spots which could cause flaring activity at all phases if in the line of sight. It is important to note here that after spectral type M4 ($\sim$ 0.3\Msun) these stars become fully convective and so do not possess a tachocline, therefore, generating their magnetic field through a different dynamo mechanism in comparison to the Sun. However, polar spots or spots with high latitudes have never been observed on the Sun, therefore, you would assume polar spots are not possible on other solar-like stars. However, in a study by \cite{schrijver2001formation} they simulate that polar spots could be possible on sun-like stars where a strong polar cap field leads to suppression of convection and formation of starspots and high latitudes. As a result, there is the possibility of polar spots being present on these solar-type stars which could be interacting with active regions at lower latitudes to produce flaring activity. Solar Orbiter \citep{muller2013solar}, an ESA mission launched in February 2020, will take high resolution images of the Sun's poles for the first time which will provide insights into the magnetic structure in more detail. 

This then brings us on to the theory of multiple spot locations. During solar maximum, the Sun can be observed to possess many active regions which host spots including multiple spots at one location. As a result, it is entirely possible for our solar-type stars to possess multiple spot locations which could produce flaring at all rotational phases. Evidence of multiple spots was observed in the one year lightcurve of HD 39150 (TIC 364588501) which was observed in the continuous viewing zone. In the \tess\ lightcurve of this particular star the shape of the rotational modulation is observed to change over time with the whole lightcurve being fold-able on one rotation period and phase zero. This suggests multiple spots which are slightly out of synchronisation producing the changes in the shape of the rotational modulation. As a result, we observe increased levels of flaring activity which co-align with the changes in the rotational modulation suggesting emergence of new spot regions and plage regions which interact with each other to produce the increased flaring activity. The large spot coverage of this and other G stars suggest youth. This is consistent with a study of the solar spectral irradiance variability over the last 4 billion years \citep{shapiro2020solarcycle}. The Total Solar Irradiance (TSI) variability of the young 600Myr old Sun was about 10 times larger than that of the present Sun with its variability been spot-dominated, while by 2.8Gyr it's variability is faculae-dominated.

As the relationship between solar flares and sunspots is well-established we used historic X-ray GOES flare data and SILSO sunspot data to investigate this relationship further. We find sunspot and flare number are closely linked across the solar cycles where sunspot numbers increase as the Sun approaches solar maximum, so does the flare activity increase. Similar to \cite{gao2016curious} we also discover the lower energy A \& B class flares are anti-correlated with the higher energy X, M \& C class. This is interesting as the lower energy solar flares are more prominent during solar minimum when the sunspot numbers are low. As a result, this suggests these solar flares could be originating from plage or local dynamo regions and are not associated or correlated with sunspots.

This leads us on to our earlier finding of starspots and stellar flares not being correlated on our sample of solar-type stars. The anti-correlation between high and low energy solar flares and the lack of a correlations between low energy flares and sunspots could aid in understanding our lack of a spot/flare connections in our solar-type stars. There is the potential for the lower energy flares, which are observed to occur at all rotational phases, to result from plage regions or local dynamo regions as well. However, this does not explain our lack of a correlations between higher energy flares and starspots as they should occur together much like what is observed on a cycle by cycle basis on the Sun. 

An alternative idea is that we should not attempt to correlate flare activity with spot number as this is not the main driver of magnetic activity. In a series of papers by \cite{mcintosh2014decipheringa, mcintosh2014magnetic} and more recently, \cite{srivastava2018extended} and \cite[and references therein]{dikpati2019triggering}; these authors suggested that activity bands belonging to the 22 year magnetic activity cycle is the main driver of solar activity, with these bands interacting at the equator. The idea behind this is an `Extended Solar Cycle' which appeared to extend the activity butterfly back in time, about 11 years before the formation of the sunspot pattern. Furthermore, these activity bands extend to much higher solar latitudes and would require a polar dynamo which is not a widely accepted idea. The observational evidence for the extended solar cycle is based on the evolution of coronal bright points, although the origin of this work dates back several years to \cite{wilson1988extended}. As noted by the authors, many large solar flares do not occur at sunspot maxima \citep[e.g. see][]{odenwald2006forecasting}. They suggest that the longer these activity bands spent at very low latitudes, the higher the probability for large flares due to the formation of complex active regions. 

With faster rotators, one may have several of these activity bands, thus a series of complex active regions producing super flares. The flare model proposed by \cite{aulanier2013standard} fails to provide sufficient energy, to explain flares on M dwarfs, e.g. if we consider a typical M3-M4 dwarf which has a radius of about 0.3\Rsun\ and a field strength of 2-4kG. From their Fig 2, these values would reproduce a flare of ~$10^{34}$~erg, however, this would require a spot to cover nearly half of the visible stellar disk and still would not be able to account for the very large flare energies on some M dwarfs. In the case of solar-type stars, \cite{aulanier2013standard} determine that a bipole coverage of 100Mm and a field strength of 4kG would be the conditions required to produce a flare of energy $10^{34}$~erg. They conclude solar-type stars which produce superflares must have stronger dynamo mechanisms than the Sun. 

This then posses the question: Would the Sun be able to produce superflares with energies > $10^{33}$~erg? In \cite{shibata2013can} they investigate this question using current ideas related to the mechanisms of the solar dynamo. In their calculations the Sun would need to generate a sunspot with magnetic flux of $2\times 10^{23}$~Mx to produce a $10^{34}$~erg flare. In order to do this it would take the Sun 40 years to store this magnetic flux and at present there is no known physical mechanism to make this possible. Overall, they conclude it is premature to say whether a $10^{35}$~erg flare would even be possible on the Sun given the current dynamo theories. 

\section{Conclusions}
We have conducted an analysis into the statistics of superflares on a sample of 209 solar-type stars. Utilising 2-min cadence data from \tess\, we derived rotation periods for our sample and characterised 1980 flares. Two of our targets were observed in the continuous viewing zone so, with one year of observations we carried out a short study into the variability of these stars. Our findings showed evidence of spot emergence, plage regions and migrating spots which were connected to increased levels of flaring activity. Overall, we focused on the relationship between rotational phase and flare number finding no correlation between the two. We included an analysis on historic solar flare and sunspot data to investigate the relationship between flares and spots on the Sun, using our results to aid in understanding the lack of a correlation in our solar-type star sample. 

By June 2020, \tess\ will finish observing the northern hemisphere and will return to sectors in the southern sky. Further observations of these stars will allow for follow up studies, investigating their changing behaviour a year later. This will provide insights into the extended magnetic activity of these stars while also allowing for the search of activity cycles. Overall, the continued study of flares and starspots can aid in understanding the dynamo mechanism of other stars and how it relates to the Sun. All of this is extremely important when considering potential habitable systems which may orbit these active host stars.

\section*{Acknowledgements}
Armagh Observatory and Planetarium is core funded by the Northern Ireland Government through the Department for Communities. LD would like to acknowledge funding from an STFC studentship. The authors would also like to the thank the anonymous referee for their comments and suggestions which helped to improve the paper. This paper includes data collected by the \tess\ mission where funding is provided by the NASA Explorer Program.
 
This work presents results from the European Space Agency (ESA) space mission {\sl Gaia}. {\sl Gaia} data is being processed by the {\sl Gaia} Data Processing and Analysis Consortium (DPAC). Funding for the DPAC is provided by national institutions, in particular the institutions participating in the {\sl Gaia} MultiLateral Agreement (MLA). The Gaia mission website is \url{https://www.cosmos.esa.int/gaia}. The Gaia archive website is \url{https://archives.esac.esa.int/gaia}. We thank the anonymous referee for their helpful report. 

The national facility capability for SkyMapper has been funded through ARC LIEF grant LE130100104 from the Australian Research Council, awarded to the University of Sydney, the Australian National University, Swinburne University of Technology, the University of Queensland, the University of Western Australia, the University of Melbourne, Curtin University of Technology, Monash University and the Australian Astronomical Observatory. SkyMapper is owned and operated by The Australian National University's Research School of Astronomy and Astrophysics. The survey data were processed and provided by the SkyMapper Team at Australian National University. The SkyMapper node of the All-Sky Virtual Observatory (ASVO) is hosted at the National Computational Infrastructure (NCI). Development and support the SkyMapper node of the ASVO has been funded in part by Astronomy Australia Limited (AAL) and the Australian Government through the Commonwealth's Education Investment Fund (EIF) and National Collaborative Research Infrastructure Strategy (NCRIS), particularly the National eResearch Collaboration Tools and Resources (NeCTAR) and the Australian National Data Service Projects (ANDS).

\bibliographystyle{mnras}
\bibliography{tess_solar.bib} 

\bsp	
\label{lastpage}
\end{document}